\documentclass[showpacs,aps,preprintnumbers,letterpaper,twocolumn]{revtex4-1}
\usepackage{amsmath,amssymb}
\usepackage{epsfig}
\usepackage{graphicx}
\usepackage{amsmath}
\usepackage{slashed}
\usepackage{amsfonts}
\usepackage{epstopdf}

\usepackage{graphicx,subfigure}
\usepackage{amsmath,amssymb}
\usepackage{hhline}
\usepackage[pdftex, pdfborder= 0 0 0, citecolor=blue, urlcolor=blue, linkcolor=blue, colorlinks=true, bookmarksopen=true]{hyperref}

\newcommand{\be}{\begin{eqnarray}}
\newcommand{\ee}{\end{eqnarray}}

\def\slashchar#1{\setbox0=\hbox{$#1$}           
   \dimen0=\wd0                                 
   \setbox1=\hbox{/} \dimen1=\wd1               
  \ifdim\dimen0>\dimen1                        
 \rlap{\hbox to \dimen0{\hfil/\hfil}}      
  #1                                        
 \else                                        
    \rlap{\hbox to \dimen1{\hfil$#1$\hfil}}   
    /                                         
 \fi}                                         %

\begin{document}

\title{Deconfinement Phase Transition in the $SU(3)$ Instanton-dyon Ensemble}

\author{ Dallas DeMartini  and Edward  Shuryak }

\affiliation{Center for Nuclear Theory, Department of Physics and Astronomy, Stony Brook University,
Stony Brook NY 11794-3800, USA}

\begin{abstract}
	Confinement remains one the most interesting and challenging nonperturbative phenomenon in non-Abelian gauge theories. Recent semiclassical (for SU(2)) and lattice (for QCD) studies have suggested that confinement arises from  interactions of statistical ensembles of instanton-dyons with the Polyakov loop. In this work, we extend studies of semiclassical ensemble of dyons to the $SU(3)$ Yang-Mills theory.  We  find that such interactions do generate the expected first-order deconfinement phase transition. The properties of the ensemble, including correlations and topological susceptibility, are studied over a range of temperatures above and below $T_c$. Additionally, the dyon ensemble is studied in the Yang-Mills theory containing an extra trace-deformation term. It is shown that such a term can cause the theory to remain confined and even retain the same topological observables at high temperatures. 
\end{abstract}
\maketitle

\section{Introduction}
\twocolumngrid

Quantum Chromodynamics (QCD) is the quantum field theory describing the fundamental particles and forces that make up nuclear physics. While QCD is remarkably successful in describing nuclear physics, many phenomena remain beyond the scope of what can be studied analytically. Notably, nonperturbative phenomena such as confinement -- the disappearance of quarks and gluons 
from the physical spectrum -- is not completely understood. Confinement occurs not just in QCD, but in various Yang-Mills theories with or without quarks, making it clear that it emerges from the non-perturbative behavior of the gluons, rather than the quarks. Above certain critical temperature $T_c$
deconfinement takes place, and the QCD-like theories turn into a new form of matter, the Quark-Gluon Plasma (QGP).  

Historically the first mechanism of the deconfinement transition 
was a 'dual superconductor' model
  \cite{Mandelstam:1974pi,Parisi:1974yh,Kondo:2014sta}.
   At $T<T_c$ the chromoelectrically-charged quarks and gluons are connected by QCD flux tubes, dual
   to magnetic flux tubes in superconductor.
With the advent of lattice gauge theories many aspects of this scenario were put to the test.
In particular, the profile of the  QCD flux tubes  \cite{Bali:1998de} was found to agree well
with dual superconductor model. Monopoles were observed and found to rotate around these
flux tubes, as expected. Bose-Einstein condensation of monopoles was detected and its 
critical temperature was shown to coincide with $T_c$ \cite{D'Alessandro:2010xg}.
A high density of monopoles was found to be responsible for unusual kinetic properties  of QGP \cite{Liao:2006ry}.

Euclidean formulation of the gauge theory lead to discovery of 4D topological solitons
known as BPST
instantons  \cite{Belavin:1975fg}.  A model of their ensemble, the Instanton Liquid Model (ILM)
\cite{Shuryak:1981ff}, has explained how instantons generate chiral symmetry breaking.
As 
 certain extrema of the path integral over gauge configurations, they form a basis for semiclassical
 theory, consistently including fluctuations around classical fields. Furthermore, 
one can study the interaction between instantons in their statistical ensemble:
those studies explained behavior of correlation functions of various  mesonic and baryonic currents, for review see e.g. Ref. \cite{Schafer:1996wv}. Yet the instanton theory has not reproduced confinement. 
 
   Euclidean formulation of finite temperature QCD naturally led to a nonzero value of the Polyakov loop $\langle P \rangle \neq 0$ as a signature of deconfinement. Note that throughout this paper we use $\langle P \rangle$ as shorthand for $\frac{1}{3} \langle  Tr[P(\vec{x})] \rangle$. It can be interpreted as the
   nonzero vacuum expectation value (VEV) of the time component
   of the gauge field $A_0$, also known as {\em nonzero holonomy}.   
  The natural question was then how to deform the instanton configurations in a way consistent with  nonzero holonomy in the bulk. It was answered 
  in Refs. \cite{Kraan:1998sn,Lee:1998bb}, who discovered that instantons dissolve  into $N_c$ (number of colors) constituent solitons, called {\em instanton-dyons} (or instanton-monopoles). Like
  original instantons, they are (anti)selfdual, so their actions and topological charges are equal. 
  But, unlike  instantons, their actions and topological charges
    are $not$ quantized to integers; standard index theorems are avoided because  instanton-dyons
    have magnetic charges and therefore are still connected by Dirac strings.
    
  It was then realized that instanton-dyons provide a very valuable bridge between the theory of monopoles and instantons, providing a way to explain $both$ confinement and chiral symmetry breaking in a single setting. Unlike monopoles, the   instanton-dyons are semiclassical objects, allowing for the construction of a consistent
   theory of an interacting ensemble. Last but not least, the statistical sum in terms of instanton-dyons 
   is "Poisson dual" to that based on monopoles, see Refs.
   \cite{Dorey:2000dt,Ramamurti:2018evz}.

   Semiclassical approaches to finite-$T$ gauge theories, with and without quarks, have lately been subject of multiple studies. Confinement in this theory is due to the back reaction of the dyon ensemble on the Polyakov loop, forcing it to take zero value at $T<T_c$, see 
   e.g. Ref. \cite{Shuryak:2013tka} for a simple model, Ref. \cite{Liu:2015ufa} for mean-field analysis, and Refs.
    \cite{Larsen:2015vaa,Lopez-Ruiz:2016bjl} for numerical simulations of the $SU(2)$ gauge theory.
         It is important
that they are semiclassical objects, unlike the QCD monopoles \cite{Ramamurti:2018evz},
with the actions $S_{dyons}\sim 1/g^2\sim log(T/\Lambda)$
   growing with temperature. Therefore their densities are suppressed at high $T$ as an inverse power of $T$. At temperatures comparable to the critical one, $T\sim T_c$, numerically $S_{dyons}/\hbar \approx 4$. The inverse of it is the small parameter of our semiclassical expansion.

    Although in this work we study pure gauge theory, rather than QCD-like theories with light quarks, let us mention
    that the instanton-dyon ensemble also describes the breaking of chiral symmetry. Numerical studies 
    of chiral phase transition can be found in Refs. \cite{Liu:2015jsa,Larsen:2015tso}.
    Using light quark Dirac eigenstates, one can identify the individual dyons
    inside lattice configurations from large-scale QCD simulations. As shown recently
     \cite{Larsen:2018crg}, the lowest Dirac states are indeed generated by the instanton-dyons,
     in the form remarkably insensitive to large density of perturbative quarks and gluons
     in which they are immersed. 
    
    In this work we extend the studies  \cite{Larsen:2015vaa,Lopez-Ruiz:2016bjl} to the pure $SU(3)$
    gauge theory. Instead of two species of instanton-dyons, we now deal with three.
    It is well-known from lattice studies \cite{Kaczmarek:2002mc} that the pure $SU(3)$ gauge theory (and all pure $SU(N_c)$ theories with $N_c \ge 3$) possesses a first-order phase transition, rather than the second-order transition seen in $SU(2)$. So the question we address below is whether this, and other, features
    of pure $SU(3)$ gauge theory can or cannot be reproduced by the  instanton-dyon ensembles.
    
    This paper is structured as follows: The dyon interactions and partition function are discussed in Section II. Section III lays out some technical details of the simulation as well as the data analysis performed. The physical results, including correlation functions and the temperature dependence of parameters such as the Polyakov loop are shown in Section IV. Finally, in Section V, the deconfinement transition of the dyon ensemble is studied in the trace-deformed Yang-Mills theory. 
      
   \section{Instanton-dyons in the $SU(3)$ gauge theory}
   \subsection{Dyons and holonomy}
   
    The Polyakov loop is an order parameter of the deconfinement phase transition
  \begin{equation}
  P = P \exp(i \oint A_{\mu}^a T^a dx_{\mu}),
  \end{equation}
  where $T^a$ is the color generator in the fundamental representation. Its VEV, $\langle P \rangle$ is a unitary matrix with phases for eigenvalues, so $A_4$ can be defined as $A_4 = 2 \pi T diag(\mu_1,\mu_2,\mu_3)$. These phases are related to the holonomies of the dyons by $\nu_i= \mu_{i+1} - \mu_i$, where $\mu_{i+1} > \mu_i$ and $\mu_4 = \mu_1$. 
  The connection between the Polyakov loop and confinement can be seen through its relation to the free energy of a static quark
  \begin{equation}
  \langle P \rangle = e^{-F_q/T}.
  \end{equation}
  Clearly, $\langle P \rangle = 0$ is the confining holonomy, corresponding to $F_q \rightarrow \infty$, removing massive quarks from the spectrum completely. 
   By combining the previous definitions of the holonomies and phases, the relationship between the holonomy and the average Polyakov line is seen to be
   \be \langle P \rangle = \frac{1}{3} + \frac{2}{3} \cos (2 \pi \nu). \ee 
   From this it is clear that $\nu=1/3$ is confining ($\langle P \rangle =0$),while $\nu\rightarrow 0$ produces the ``trivial" Polyakov loop ($\langle P \rangle =1$). 
   
 For $SU(N_c)$ gauge theories, there are $N_c$ types of dyons: $N_c-1$ of them are called $M_i$-type dyons, they correspond to the maximal diagonal subgroup. One more type is called the
  "twisted" (that is, 4-time-dependent) or $L$-type dyon. There are corresponding antidyons as well giving in total $2N_c$ species of dyons. Note that we will use \textit{species} to refer to one of the six dyons and \textit{type} to refer to one of the three pairs of a dyon and its antidyon (i.e. an $M_1$-type dyon refers to both $M_1$ and $\bar{M}_1$ dyon species).  The action of each individual dyon of type $i$ is denoted by $S_0 \nu_i$, where $\nu_i$ are the so-called holonomy parameters. Those are also  called ``fractions of the holonomy circle"
  because they satisfy the sum rule
   $$\sum_i^{N_c} \nu_i = 1$$ and {\em generically} have $N_c-1$ independent parameters. These holonomy parameters determine not only the actions 
 but also the core sizes of the dyons $r_i \sim 1/(2\pi \nu_i T)$. In $SU(N_c)$ theories, (anti)instantons are comprised of one of each of the types of (anti)dyons, although this is not necessarily the cause for other gauge groups. For example, in $SU(3)$, the instanton $I = M_1 L M_2$ and the antiinstanton $\bar{I} = \bar{M_1} \bar{L} \bar{M_2}$.
 
 The main parameter defining the properties of the dyons is the classical instanton action
  $$S_0 = {8\pi^2\over g^2}\sim log(T)$$
 containing the coupling at the temperature-dependent scale. 
  While this parameter is independent of the holonomy, the actions of individual dyons of type $i$ are $S_i = S_0 \nu_i$. In the $SU(2)$ case, one defines one holonomy parameter $\nu\in [0,1]$, the $M$ dyon having action $S_M=S_0 \nu$ and   
  the $L$ dyon having the conjugate action $S_L= S_0 (1-\nu)$. 
  In the $SU(3)$ case  there are two diagonal Gell-Mann matrices, $\lambda^3$ and $\lambda^8$ and in principle two holonomy parameters. However, the average Polyakov loop $\langle P \rangle$ is a gauge-invariant, and thus physical, object. Restricting it to be $real$ enforces $\nu_{M1} = \nu_{M2}$ reducing the holonomy to just a single parameter for $SU(3)$  with $\nu\in [0,1/2]$. The dyon actions expressed in it are 
   \be  S_{M1}= S_{M2} =S_0 \nu, \,\,\,\,\, S_L= S_0 (1-2\nu). \ee
  This symmetry between the $M_1$- and $M_2$-type dyons means that they have equal densities as well, greatly reducing the space of parameters that needs to be considered.  
  
  \begin{figure}[h]
  	\includegraphics[width=7cm]{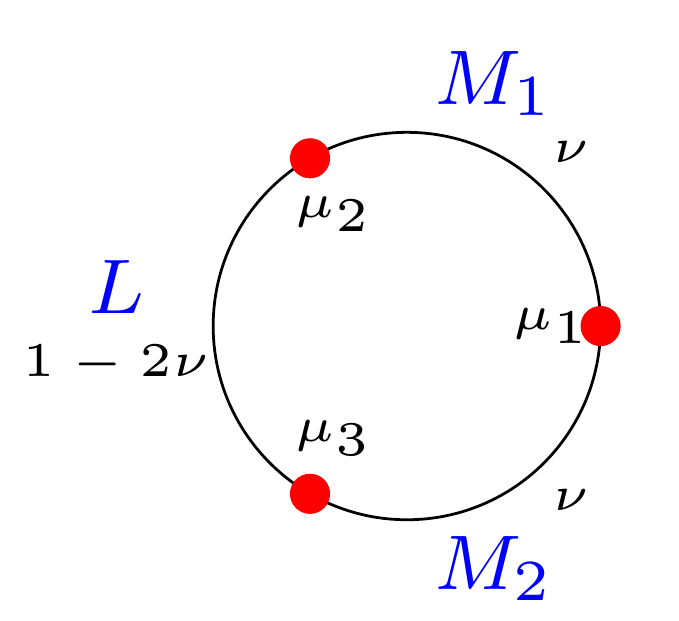}
  	\caption{(Color online) Structure of the holonomies and dyon types in $SU(3)$. Circle has circumference of 1.}
  \end{figure}

   \subsection{The partition function and dyon interactions}
   A complete calculation of the dyons' free energy requires the construction of the dyonic partition function. We start first with effects that are not induced by the dyons' non-perturbative interactions. In the absence of all dyonic effects, there is a perturbative interaction between thermal gluons and the holonomy. This generates the Gross-Pisarski-Yaffe potential $V_{GPY}$ \cite{Gross:1980br}, which appears in the an exponent in the partition function. For $SU(3)$, this is given by 
   \begin{equation}
   \frac{V_{GPY}}{T\tilde{V}_3} = \frac{4 \pi ^2}{3}(2(\nu(1-\nu))^2 + (2\nu(1-2\nu))^2).
   \end{equation}
   Here the potential is shown with factors divided out so that it has units of free energy density. This potential disfavors confinement, having a minimum at the trivial holonomy $\nu=0$, and a maximum at the confining holonomy $\nu =\frac{1}{3}$. 
   
   The dyon contribution to the partition function 
   consists of two parts,  the contributions of the dyons in the absence of interactions $Z_0$, which can be expressed analytically directly from the input parameters, and $Z_{np}$ the contributions of the dyon interaction, whose calculation is the subject of the simulation of the partition function,  performed  by Monte-Carlo algorithms in this work. The total partition function is their product $Z = Z_0 Z_{np}$.
   
   The statistical weight for a single instanton was explicitly calculated in Ref. \cite{Diakonov:2004jn} for $SU(2)$. Taking the dilute limit, which removes interaction between the dyons, this gives 
   \begin{equation}
   Z_0^{SU(2)} = \frac{\Lambda}{(4 \pi)^2} S_0^4 e^{-S_0^2} \nu^{\frac{8\nu}{3}-1} (1-\nu)^{\frac{8(1-\nu)}{3}-1}
   \end{equation}
   It is easy to see how this factors into the weight for each individual dyon: Each dyon species contributes a factor of $ \sqrt{\Lambda}S_0 e^{-\sqrt{S_o \nu}}$ and the two holonomy terms stem from the holonomies of the individual M- and L-type dyons. A factor of $4\pi \nu_i$ is divided out for each dyon type. This is to remove a constant term that appears remains when taking the dilute limit in $Z_{np}$. Knowing this, it is easy to construct the instanton weight in $SU(3)$ and extend it to an arbitrary number of dyons by summation. The partition function, assuming an equal number of dyons and antidyons, is
   \begin{equation}
   \begin{aligned}
   Z_0= &\sum_{N_{M1},N_L,N_{M2}} \left(\frac{1}{N_{M1}!} (\tilde{V}_3 d_{\nu})^{N_{M1}} \right)^2 \\
   & \times \left(\frac{1}{N_L!} (\tilde{V}_3 d_{1-2\nu})^{N_L} \right)^2 \\
   & \times \left(\frac{1}{N_{M2}!} (\tilde{V}_3 d_{\nu})^{N_{M2}} \right)^2 ,
   \end{aligned}
   \end{equation}
   where $d_{\nu}$ is the weight of an individual dyon with holonomy $\nu$,
   \begin{equation}
   d_{\nu} = \frac{\Lambda}{4\pi} S_0^2 e^{-S_0 \nu} \nu^{\frac{8 \nu}{3} - 1}
   \label{weight}.
   \end{equation}
   
   Now we may take the limit $V \rightarrow \infty$ and assume equal densities of $M_1$- and $M_2$-type dyons. Additionally we resolve the factorial terms with Stirling's approximation carried out to three terms $\ln N! \approx N \ln N - N + \ln (\sqrt{2\pi N})$. The free energy $F = -\ln(Z)$ is given by the following expression    \begin{equation}
   \begin{aligned}
   f ={}& \frac{4 \pi ^2}{3}(2(\nu(1-\nu))^2 + (2\nu(1-2\nu))^2) \\
   & -4n_M \ln \left[\frac{d_{\nu}e}{n_M}\right]- 2n_L \ln \left[\frac{d_{1-2\nu}e}{n_L}\right] \\ 
   & + \frac{\ln(8\pi^3 N_M^2 N_L)}{\tilde{V}_3} +\Delta f \\
   \end{aligned}
   \label{eqfree}
   \end{equation}
   where $\Delta f$ is the free energy density stemming from the interactions of the dyons. If the dyons have classical binary interactions $\Delta S_{class}$ and a volume metric $G$, their contributions to the partition function and the free energy density are 
   \begin{equation}
   Z_{np} = \frac{1}{\tilde{V}_3^{(4N_{M1}+2N_L)}} \int Dx \det{(G)} e^{-\Delta S_{class}}
   \end{equation}
   \begin{equation}
   \Delta f = -\ln(Z_{np})
   \end{equation}
   The set of parameters that minimizes the free energy density  corresponds to the physical  dyon ensemble in the infinite volume limit. This elucidates the main procedure of this work: to first compute the free energy density of the ensemble for a wide range of parameters,  and then locate
   their values  which minimize it.
   
   The classical interactions between dyons and antidyons are, at distances exceeding the dyon cores, asymptotically Coulomb-like. For generic $SU(N_c)$ theories, the interactions are given in Ref. \cite{Unsal:2008ch}. At shorter length scales, the interaction is modified. The interaction between dyons and antidyons of the same type was studied in detail and parameterized by Larsen and Shuryak \cite{Larsen:2014yya}. In the previous $SU(2)$ model, this interaction was used for dyon-antidyon pairs of the same type, while dyons and antidyons of different types had purely Coulomb-like interactions. In this work we use a single modified parameterization for all dyon-antidyon pairs. The parameterization used is given by
   \begin{equation}
   \Delta S_{class}^{d \bar{d}}= -\frac{S_0 C_{d \bar{d}}}{2\pi}(\frac{1}{rT}-2.75 \pi \sqrt{\nu_i \nu_j} e^{-1.408 \pi \sqrt{\nu_i \nu_j} r T}),
   \end{equation}
   where $C_{d \bar{d}}$ is a coefficient with value $2$ for pairs of the same type and $-1$ for pairs of different types. Compared with the parametrization used in $SU(2)$, two modifications have been made: the substitution $\nu_i \rightarrow \sqrt{\nu_i \nu_j}$ was made to accommodate pairs of dyons with different holonomies (e.g. $M_1 \bar{L}$), and the coefficient in front of the exponential term has been reduced from the original values of $3.264$.
   These interactions are used for distances greater than the core size $r_0$ = $x_0/(2 \pi \nu T)$. Dyons pairs of the same type (regardless of duality) experience a repulsive core. While the core potential has not been studied in detail it reasonably described by
   \begin{equation}
   \Delta S_{class}^{core} = \frac{\nu V_0}{1+e^{2 \pi \nu  T (r - r_0)}}.
   \end{equation} 
   The two parameters in this interaction, $x_0$ and $V_0$ are chosen on a phenomenological basis and should be subject to constraints from appropriate lattice data when possible. See Appendix B for a discussion of their values and the effects of changing them. 
   
   Additionally, the dyons experience an effective potential from the fluctuation determinant of the instanton. This effect leads to the Diakonov determinant \cite{Diakonov:2009jq} of the metric of the space of dyons' collective variables
   \begin{equation}
   \begin{aligned}
   G_{im,jn} ={} & \delta_{ij} \delta_{mn} (4\pi \nu _m - \sum_{k\ne i} \frac{2}{T|r_{i,m} - r_{k,m}|}\\
   & + \sum_{k} \frac{1}{T|r_{i,m}-r_{k,p \ne m}|}) \\
   &+ \frac{2\delta _{mn}}{T|r_{i,m}-r_{j,n}|}-\frac{\delta_{m \ne n}}{T|r_{i,m}-r_{j,n}|},
   \end{aligned}
   \end{equation}
   where $r_{i,m}$ is the position of the i'th dyon of type m. This metric only accounts for the selfdual dyons. An equivalent metric $\bar{G}$ is used for the antidyons as well. The metric is true for dyons of different species at any distance, but only true at larger distances for dyons of the same species. We modify the terms with a cutoff $r\rightarrow \sqrt{r^2 + (3/2\pi T)^2}$ so that the diagonal entries go to $0$ rather than $-\infty$ when $\nu = \frac{1}{3}$. The effective potential from this metric comes as 
   \be \Delta S_{1-loop} = -\ln (\det G \det \bar{G}). \ee
   The diagonal of this metric does not vanish in the dilute limit, instead going to $\prod_i 4 \pi \nu_i$. This has been accounted for by modifying the individual dyon weight in Eq. (\ref{weight}).
   
   It is observed that in the case the densities of different dyon types are unequal, then in the infinite volume limit, the sum diverges. We therefore regulate all Coulomb terms in both the classical and one-loop potentials with the dimensionless Debye mass $r \rightarrow re^{M_D r T}$. Like the core parameters, the value of the Debye mass is a choice that should be subject to improvement. 
   
   Now let us give a brief qualitative discussion of how such interactions should generate confinement. The main interaction driving confinement is the repulsive cores of the dyons. Because the volume of the cores goes as $1/\nu^3_i$, this interaction disfavors any type of dyon becoming large and thus favors the confined phase $\nu = \frac{1}{3}$ where all dyons are the same size. The long-distance classical Coulomb interactions disfavor confinement however. Because dyons repel anti-dyons of different types but attract antidyons of the same type, these interactions favor ensembles with many dyons of the same kind - a large value of $n_M/n_L$ - rather than equal numbers of all species. At larger values of $n_M/n_L$ the entropy portion of the partition function $Z_0$ more strongly favors small values of $\nu$, driving the system to the deconfined phase. The one-loop Coulomb-like interactions have the opposite sign and thus opposite effect. These one-loop interactions are suppressed by a factor of $S_0$ and are only comparable to the classical interactions at low $T$.
   
   The nonperturbative interactions grow weak compared to the perturbative portions of the partition function as $T$ increases. At low temperature we have an ensemble where the cores and one-loop interactions dominate, binding the dyons into instantons and at high temperatures we have a gas of mostly $M_i \bar M_i$ pairs bound by their classical attraction and the entropy and Gross-Pisarski-Yaffe terms have shifted $\nu$ to a lower value. 

   \section{The Simulation Setting and Data Analysis}

Numerical integration over the dyons' coordinates is performed by  Monte-Carlo techniques. The position of each dyon is updated and then accepted or rejected according to the standard Metropolis algorithm. Once this is done for each dyon five times (five updates is approximately the autocorrelation time of the ensemble), the configuration is sampled and its properties are computed. The free energy is obtained via the usual integration over auxiliary parameter $\lambda$:
\begin{equation}
e^{-F(\lambda)/T} = \int Dr e^{-\lambda S},
\end{equation} 
\begin{equation}
\Delta F = \int_0^1 d\lambda T \langle \Delta S \rangle,
\end{equation}
where this nonperturbative free energy is added to the perturbative contribution from Eq. \ref{eqfree}. Integration over the dummy parameter is performed in 10 equal steps $\lambda= 0.1$, ... $1$, with the average energy at each step being computed from 2000 configurations. Because the contribution at small $\lambda$ is large, the value of the average interaction at $\lambda = 0$ is obtained from a linear fit of the nearest points and included in the numerical integration. With all of this, the statistical uncertainty in the computations of $f$ are generally at the percent level. 

The goal of these simulations is to calculate the free energy density $f$ of ensembles with three main input parameters, the holonomy $\nu$ and the dyon densities $n_M$, $n_L$ over a range of temperatures near $T_c$. All simulations are performed with a fixed number of dyons $N_D = 120$ or $122$ (60 or 61 dyons and 60 or 61 antidyons). For the six species of dyons in $SU(3)$, $N_D = 4 N_M + 2N_L$. The system is studied on flat geometry with finite density imposed by a periodic box surrounded by 26 image boxes, as was done for $SU(2)$ in Ref. \cite{Lopez-Ruiz:2016bjl}. The densities of the dyons, $n_M$ and $n_L$ are controlled by varying the size of the box and the ratio of the numbers of each type of dyon $N_M/N_L$ for a specific value of $N_D$. Two values of $N_D$ are used to allow for more values of $N_M/N_L$ to be used.   

There are two technical details of the simulation that must be mentioned. The first is the Diakonov determinant $\det G$, which 
is not positive definite for randomly-placed dyons. It has been shown by Bruckmann et. al. \cite{Bruckmann:2009nw} that in fact, at reasonable densities, nearly all random configurations have at least one negative eigenvalue $\lambda_i$. This is handled by 
a modification of the effective potential
\begin{equation}
\Delta S_{1-loop} = \left\{
\begin{array}{ll}
-\ln(\det G \det \bar{G}) \quad & \text{if} \, \lambda_i \geq 0 \,\,\, \forall \, i \\
V_{max} \quad & \text{else} \\
\end{array}
\right.
\end{equation} 
 where $V_{max}$ is a large value (we choose $V_{max} = 100$) which serves to reject all Metropolis updates into the region of negative eigenvalues. The choice of $V_{max}$ is arbitrary so long at it is large enough to ensure that no configuration with negative eigenvalues is accepted during the course of the simulation.
 
 \begin{figure}[h]
 	\includegraphics[width=8.5cm]{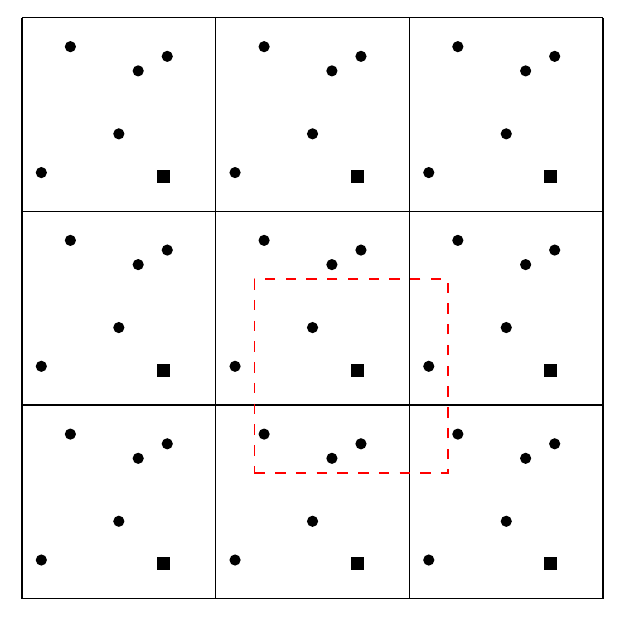}	
 	\caption{(Color online) Schematic depiction of the main simulation box (center) and surrounding images setup. Dashed box shows the volume containing all dyons whose interactions are considered in the Metropolis update step of the dyon represented by a square point. Such a volume always contains $N_D$ dyons.}
		 \label{fig_box}
 \end{figure}
 
 The second is an approximation which must be made due to computational resources. The total system, the main box and all its nearest images, contains 27$N_D$ dyons.
 Its 2D version is shown in Fig. \ref{fig_box}. The most demanding step in the simulation is the computation of $\det G$ which is an $\mathcal{O}(N^3)$ process. Performing this computation at every update step of the Metropolis algorithm is numerically costly, given the large number of parameters for which the free energy must be calculated. Instead, the change in action associated with each update is computed considering only the dyons in a box of the same size as the simulation box centered at the position of the dyon being updated. This approximation is well-justified by the fact that all long-range interactions are regulated by a Debye screening length.

Once the free energy density of each set of parameters is determined, a fit is performed to determine the dependence of the free energy density near the minimum. For each value of the action $S_0$ there is a 3-dimensional space of input parameters with coordinates $(\nu, n_M, n_L)$. The points near a minimum are fit to a quadratic function in these coordinates. From the fit, the minimum value of $f$ and its location in the coordinate space is determined. For every value of $S_0$ tested, all local minima are identified and the points near them are fit as described. The fit with the lowest minimum value of $f$ is the global minimum. For cases where a minimum is located in the in the confined phase with $\nu = \frac{1}{3}$ and $n_M=n_L$, these equalities are assumed exact and the minimum is fit to a quadratic function only in $n_M$.

    \begin{table}[h]
    	\caption{Input parameters used for the main simulation runs. Some additions and changes were made as needed. (Some non-integer values of $S_0$ were added near $T_c$ and the values of $n_M$ were increased at lower temperatures.)}
    	\begin{tabular}{|c|c|c|c|c|}
    		\hline
    		& min. & max. & step size & no. of steps \\
    		\hline
    		$S_0$	& 8 & 21 & 1 & 14 \\
    		\hline
    		$\nu$ & 0.1 & 0.35 & $0.01\bar{6}$ & 16 \\
    		\hline
    		$n_M$	& 0.15 & 0.6 & 0.03 & 16 \\
    		\hline
    		$N_M/N_L$ & 1.0 & 30.0 & varied & 16\\
    		\hline	
    	\end{tabular}
    	\label{tabparam}
    \end{table}

In summary, the results presented in this work are obtained by the following procedure: 
\\(i) perform simulations over a wide range of parameters $\nu$, $n_M$, $n_L$ for each temperature tested, \\(ii) determine the global minimum for each temperature via fitting to determine the physical values of these parameters $\nu(T)$, $n_M(T)$, $n_L(T)$, \\(iii) study the ensembles with the physical values with increased statistics to compute correlations and topological charge distributions. \\ (iv) With these increased-statistics runs, we generate 80,000 configurations with the physical interactions between them ($\lambda=1$, we no longer need to integrate over it) using the fitted values of the inputs for each temperature studied. 

\onecolumngrid  
  
      \begin{figure}[t]
      	\includegraphics[width=0.9\linewidth]{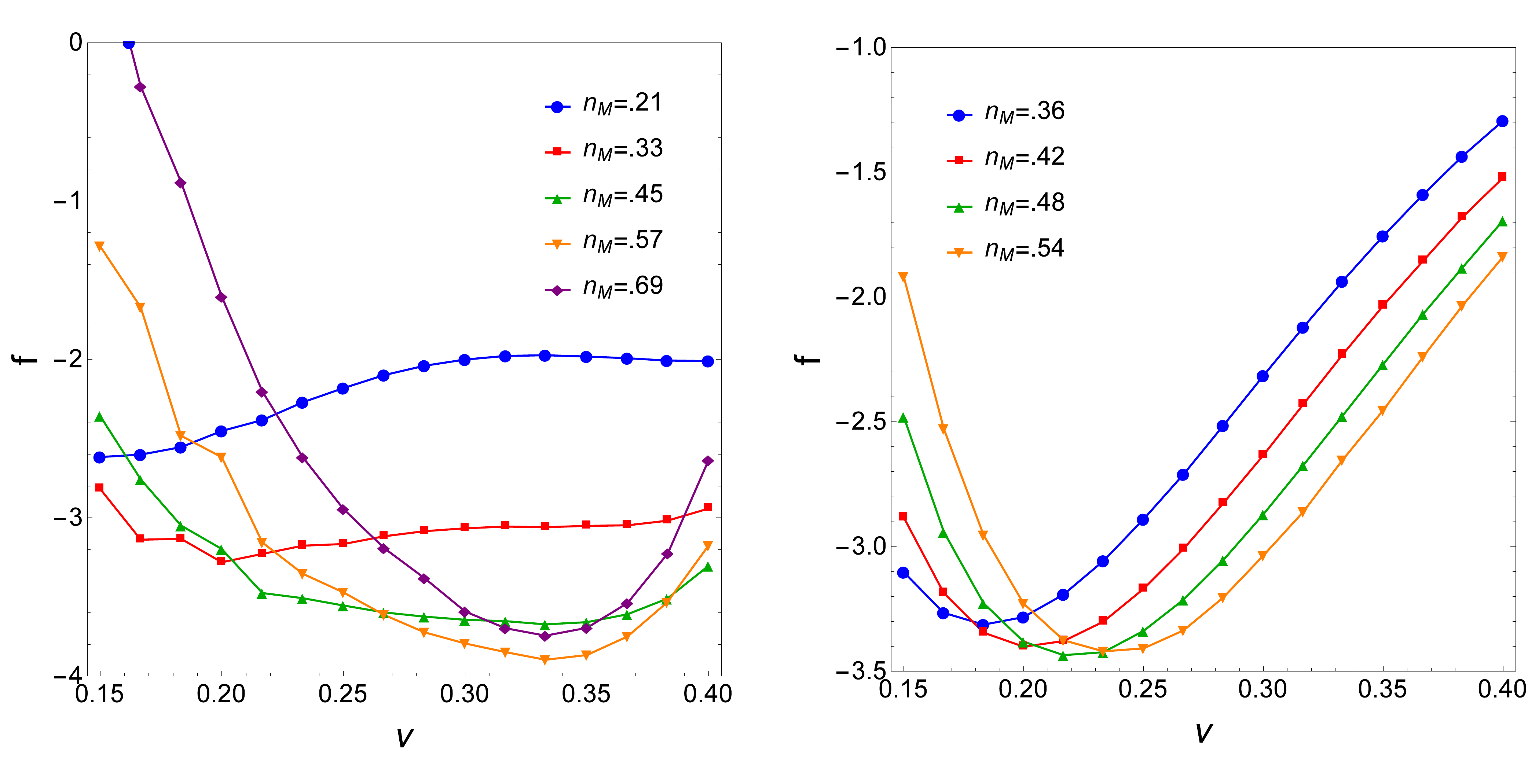}
      	\caption{(Color online) Holonomy dependence of the free energy density for different values of the $M$-dyon density $n_M$ in both phases. Left: confined phase with $S_0=12.75$, $n_L=n_M$ Right: deconfined phase with $S_0=13.75$, $n_L=n_M/9.\bar{6}$. Error bars not shown for readability.}
      	\label{fig_struc}
      \end{figure}
   
\twocolumngrid

   \section{Results} 
   
   \subsection{Structure of the holonomy potential and the phase transition}

   It is expected that the pure $SU(3)$ gauge theory possesses a first-order phase transition at some critical temperature $T_c$. The dyons interactions generate the holonomy potential which has a global minimum in the confined phase at $T < T_c$ and then a jump to a new global minimum in the deconfined phase at $T > T_c$. The first goal of this work was to check for the existence of such a phase transition. Parameters of the interaction were chosen so this occurs around $S_0 \sim 12$ (See Appendix B for more details). For each value of the temperature, which is related to $S_0$ by Eq. (\ref{eqtemp}), the shape of the holonomy potential was computed and the minimum was determined. The value of the holonomy and dyon densities at which this minimum occurs are the physical values the ensemble takes on at that temperature. 
   
   The instanton action $S_0$ was varied in steps of 1 from $S_0 = 8$ - $21$. It was observed that the phase transition occurs between $13<S_0<14$ and additional simulations were performed at $S_0 = 12.75$, $13.25$, $13.5$, and $13.75$ to locate the critical temperature more accurately. A linear fit was performed to the free energy density with the nearest points on both sides of the transition to determine  that the phase boundary is located at $S_0 = 13.18$. With this, the action can now be rewritten in terms of $T/T_c$. The range of temperatures studied here is then $0.62T_c < T < 2.04T_c$. In physical units, this is approximately $160$ MeV $< T < 530$ MeV.
   
   This model of the dyon interactions relies on a semiclassical expansion. An important consideration is then, whether or not the main semiclassical quantity, the action of the dyons $S_i = S_0 \nu_i$, remains sufficiently large over this temperature range. In the confined phase, all dyons have action $S_0/3$, meaning there is clearly a lower bound where the action becomes small and we choose $S_0=8$, $S_i=2.\bar{6}$ as a reasonable lower bound. In the deconfined phase, as the instanton action grows, the holonomy decreases making the action of the $M_i$-type dyons the limiting factor. The action of these dyons remains nearly constant at $S_M \simeq 3$ up to $2T_c$. Thus, the dyons remain sufficiently semiclassical throughout the entire range of temperatures studied. 
   
    \begin{table}[h]
    	\caption{Values of the parameters of the ensemble above and below the critical temperature $T_c$ from linear fits to the nearest data points on either side of the phase transition at $S_0 =13.18$.}
    	\begin{tabular}{|c|c|c|}
    		\hline
    		 & $T \rightarrow T_c^-$ & $T \rightarrow T_c^+$ \\
    		\hline
    		$\nu$	& $1/3$ & 0.236 \\
    		\hline
    		$\langle P \rangle$ & 0 & 0.392 \\
    		\hline
    		$n_M$	& 0.550 & 0.529 \\
    		\hline
    		$n_L$	& 0.550 & 0.068 \\
    		\hline	
    	\end{tabular}
    	\label{tabphase}
    \end{table}
   
   By performing similar linear fits, all parameters of the ensemble can be determined on both sides of the phase transition. Table \ref{tabphase} summarizes these values. From this table it is clear that the transition is first order - the holonomy / Polyakov loop and both dyon densities are discontinuous at the phase boundary.
   
    At temperatures below $T_c$ the ensemble is in the confined phase with $\nu= 1/3$ and $n_M= n_L$, as can be seen in Fig. \ref{fig_struc} (left). At densities below the physical one, the minimum shifts to the left as the nonperturbative interactions become weak compared to the perturbative contribution to the free energy. At higher densities the ensemble prefers to remain in the confined phase, but with a larger free energy minimum and curvature of the potential. 
    
    The deconfined phase has a similar structure, but with the global minimum occurring at $\nu < 1/3$ and $n_M > n_L$. However at densities above that of the global minimum, the minima continue to move towards larger $\nu$. At these densities, the repulsive core dominates and it becomes energetically favorable to make the many $M_i$ dyons smaller at the cost of making the few $L$ dyons larger. It is possible that at densities higher than what were studied here, the ensemble may have a minimum at $\nu > 1/3$.
    
\begin{figure}[h]
       	\includegraphics[width=0.9\linewidth]{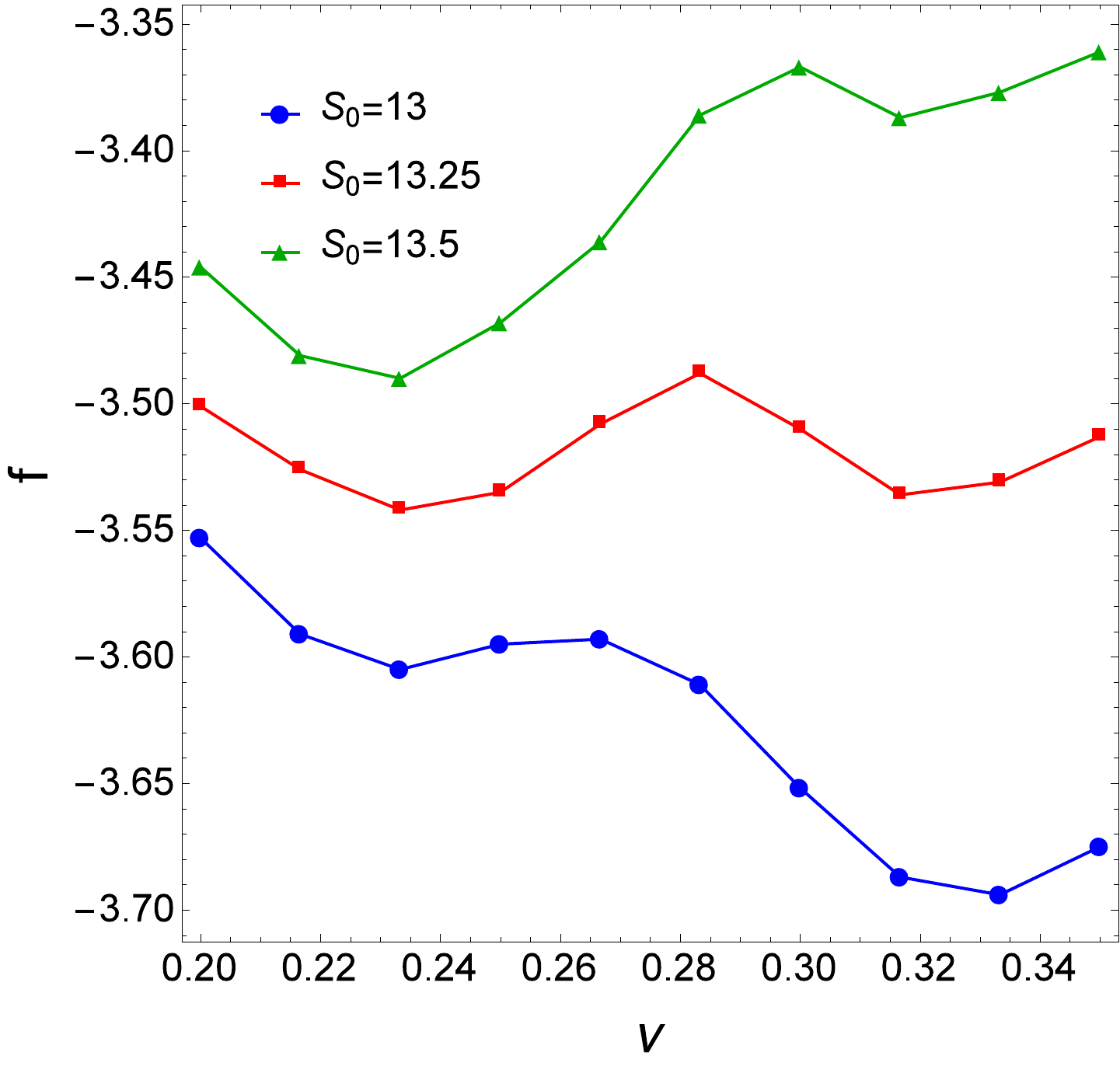}
       	\caption{(Color online) Holonomy dependence of the minimum free energy density near the phase transition. Error bars not shown for readability.}
       	\label{fig_struct2}
       \end{figure}

    These plots only show a slice of the full space of parameters explored for each value of $S_0$ for specific values of $n_M/n_L$. The structure of the first-order phase transition can be seen more clearly in Fig. \ref{fig_struct2}. By considering the minimum free energy density selected from all combinations of $n_M$ and $n_L$ as a function of the holonomy, the two local minima -- one in the confined phase and the other in the deconfined -- are clearly visible. At the value of $S_0$ nearest to the critical value, the free energy at the two minima are nearly degenerate and the global minimum switches between the two as the temperature changes. 
    
    This structure is different from the that of $SU(2)$. In $SU(2)$, where the phase transition is second order, rather than having two degenerate minima, the holonomy potential flattens near $T_c$ (see e.g. Fig. 5 of Ref. \cite{Larsen:2015vaa}). This allows the minimum to quickly, but smoothly, shift from the confining holonomy to smaller values. Additionally, there is a $\nu \leftrightarrow 1 - \nu $ symmetry not present in $SU(3)$.
   
   \subsection{Temperature dependence of the parameters}
   
\begin{figure}[h]
   	\includegraphics[width=0.9\linewidth]{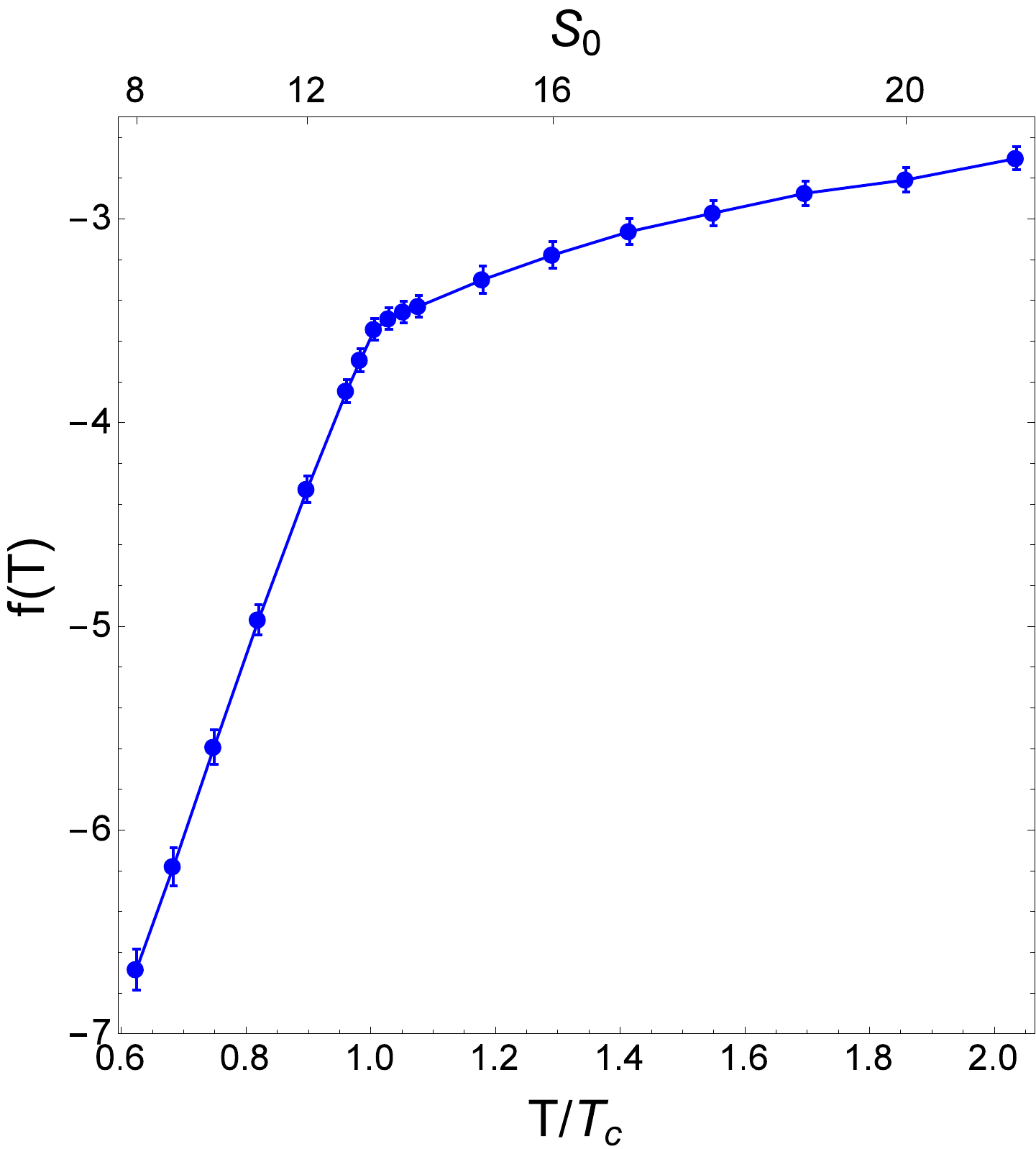}
   	\caption{Temperature dependence of the free energy density of the dyon ensemble.}
   	\label{fig_free}
   \end{figure}

   The free energy density $f$, unlike other physical quantities, remains continuous across even a first-order phase transition, as we see in Fig. \ref{fig_free}. Its derivative, however, may not. The free energy varies with temperature much more rapidly in the confined phase than the deconfined.  
   
	The most important feature of the dyon ensemble for describing the deconfinement transition is the average Polyakov loop as a function of the temperature $\langle P(T) \rangle $. Below $T_c$, the holonomy takes the confining value $\nu = 1/3$, $\langle P \rangle = 0$. At $T_c$ the value jumps to $\sim 0.4$ and then continues to increase as $T$ increases. The value of the average Polyakov loop above the phase transition shows qualitative agreement with the lattice data \cite{Kaczmarek:2002mc}, but does not increase with temperature as quickly. A change to the parameters of the dyon interactions could improve the agreement with the lattice data.

   \begin{figure}[t]
    \includegraphics[width=0.9\linewidth]{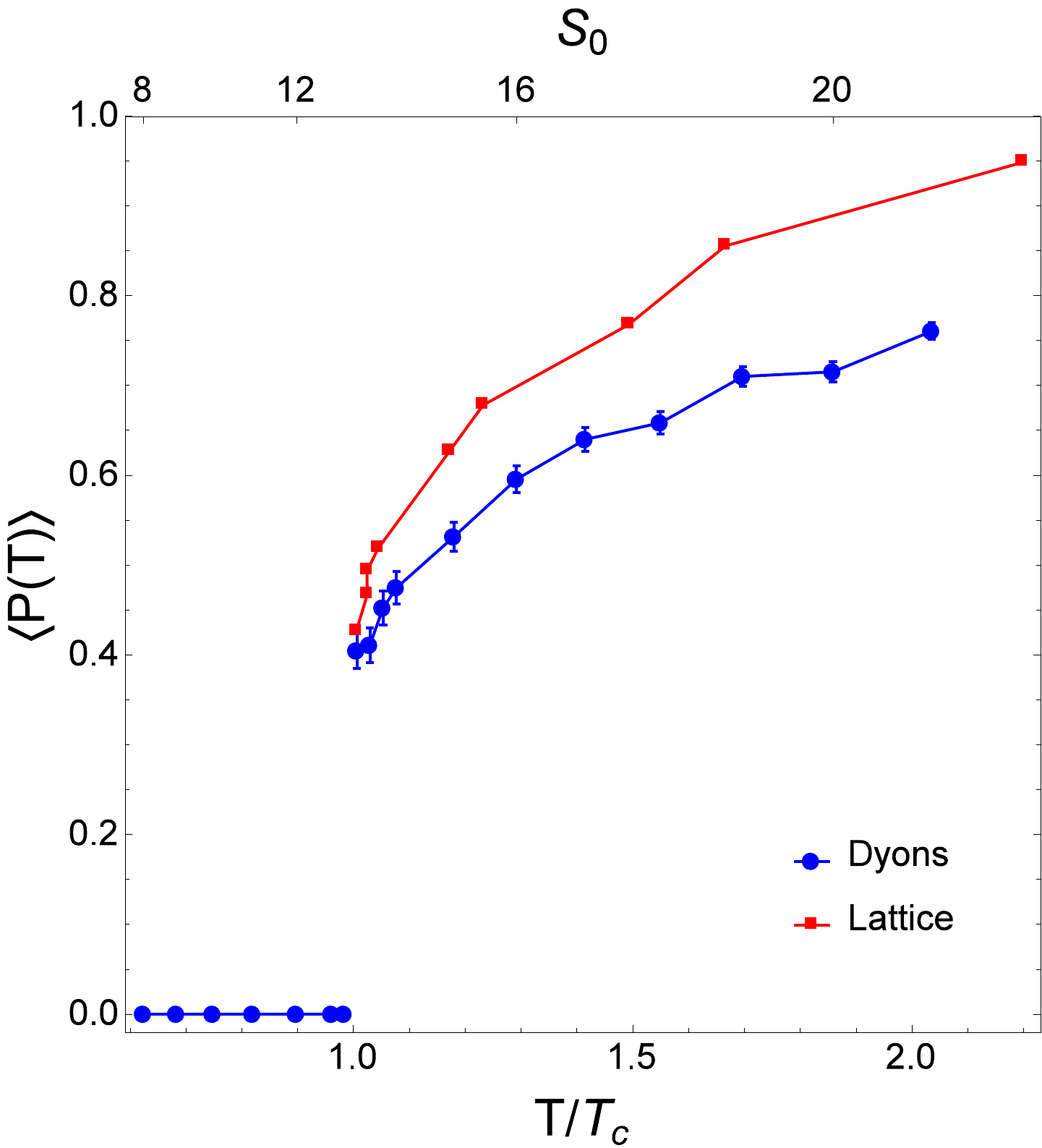}
    \caption{(Color online) Temperature dependence of the average Polyakov loop of the dyon ensemble. Lattice data taken from Ref. \cite{Kaczmarek:2002mc} and shown without error bars. Error on lattice data has magnitudes comparable to dyon data.}
    \label{fig_poly}
   \end{figure}

\begin{figure}[t]
    \includegraphics[width=0.9\linewidth]{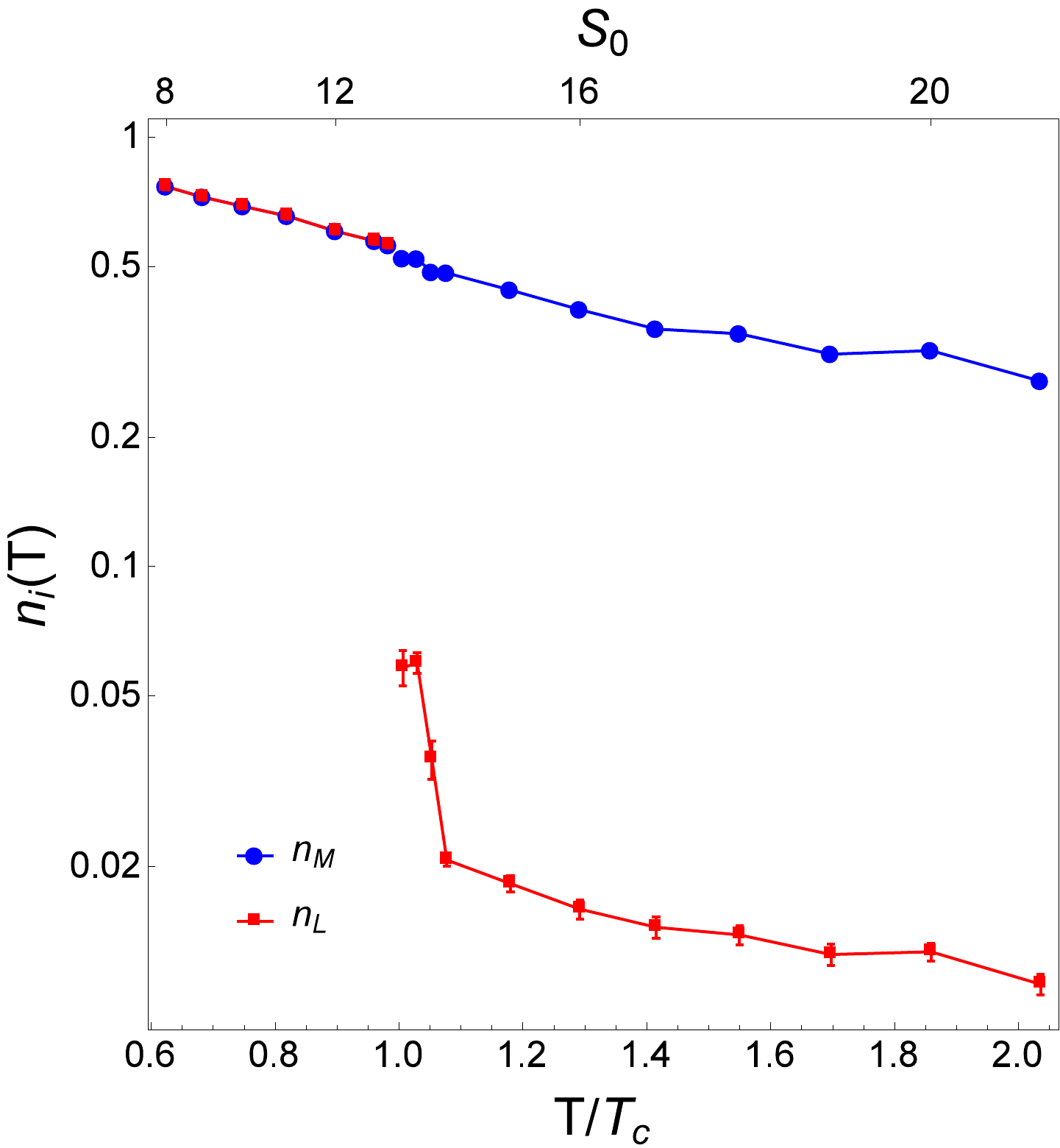}
    \caption{(Color online) Temperature dependence of the densities of each type of dyon in the ensemble.}
    \label{fig_dens}
   \end{figure}

    Another set of properties of the ensemble are the densities of each dyon type, shown in Fig. \ref{fig_dens}. At $T < T_c$, all densities are equal reflecting the fact that, in the confined phase, all dyons have equal statistical weights, core sizes, and masses and have symmetric interactions between them. At $T > T_c$, the holonomy decreases discontinuously, causing a similar change in both dyon densities. In the case of the $M$-dyon density $n_M$, the decrease in $\nu$ increases the statistical weight of the $M_i$-type dyons, while simultaneously increasing the size of their repulsive cores. These competing effects result in what is seen in the ensemble: a small, $\mathcal{O}(5\%)$, decrease in $n_M$ at the phase transition. The $L$-dyon density sees a more significant decrease due to the large decrease in statistical weight of the $L$ dyons from the increase in $1-2\nu$. 

    Overall, the dyon densities decrease with temperature, consistent with the expected behavior of instantons. It is the dyon densities that set the upper limit on the temperature that can be studied in our ensemble. As temperature increases, the ratio $n_M/n_L$ becomes larger and larger, and, at the largest values, the number of $L$ dyons in the simulations is just 1 or 2. The small number of $L$ dyons makes fitting near the free energy minimum in $n_L$ difficult as their contribution to $f$ is vanishingly small. Thus, our upper limit ($S_0=21$) is a technical constraint, rather than a physical one; accurately probing higher $S_0$ would require a larger ensemble with better statistics.
   
   \subsection{Spatial correlations between dyons}
   
   It is useful to study the effects of the dyons' interactions by studying the spatial correlations between them. Such correlations will be useful for comparison to studies of the dyons in lattice configurations. The most straightforward characteristic of the spatial correlations are the correlation functions $C_{ij}(rT)$ between two dyons of species $i$ and $j$. In $SU(3)$, because the instanton is comprised of three constituent dyons, it is useful to also define the 'instanton correlation function' $C_I(yT)$, where $y$ is the hyperdistance in the $6$-dimensional space of Jacobi coordinates of the three dyons
   \begin{equation}
   y^2 = \frac{1}{3} \left( r^2_{M1,L} + r^2_{M1,M2} + r^2_{L,M2} \right).
   \end{equation}
   In both cases, the functions are normalized such that $C_{ij}$, $C_I \rightarrow 1$ at large (hyper)distance and have had the angular factors divided out ($(rT)^2$ for the two-dyon correlation functions and $(yT)^5$ for the instanton correlation function).
   
\begin{figure}[t]
    	\includegraphics[width=0.9\linewidth]{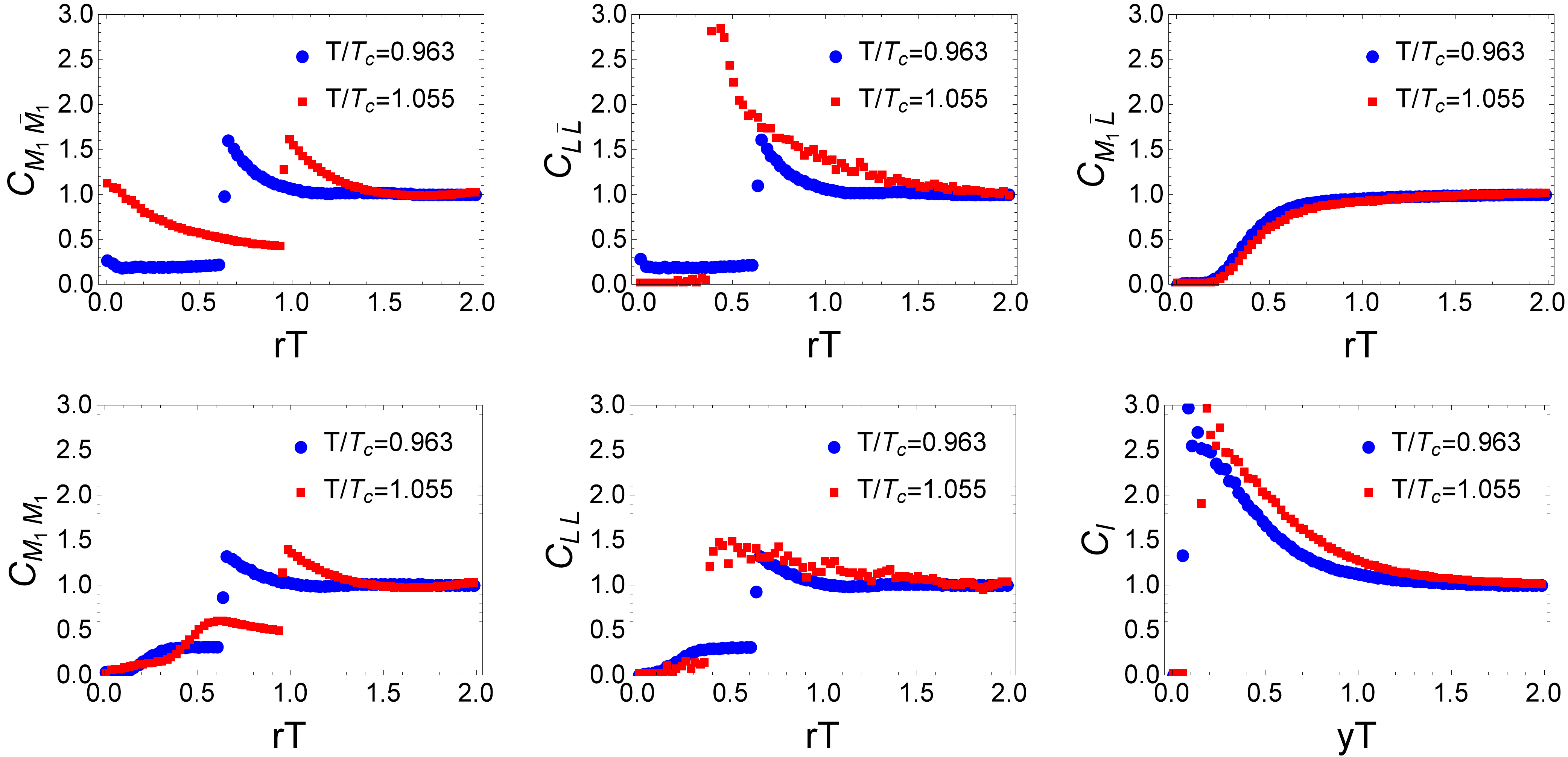}
    	\caption{(Color online) Spatial correlations of the dyons in the $M_1 \bar{M_1}$ (upper right), $L \bar{L}$ (upper middle), $M_1 \bar{L}$ (upper right), $M_1 M_1$ (lower left), $L L$ (lower middle), and instanton (lower right) channels above and below the phase transition.}
    	\label{fig_corr}
    \end{figure}

   Many of the correlation functions are redundant so we need to only observe a subset of the correlation functions to understand the behavior of the ensemble. Fig. \ref{fig_corr} shows these functions in both phases. The strongest correlation between the dyons is seen in the instanton channel where the three constituent dyons have positive short-range correlation due to the attractive terms in the Diakonov determinant.
  \newpage 
    \onecolumngrid
    
    \twocolumngrid 
       
    In several channels we can see clearly the effects of the repulsive core at short ranges. In the $D\bar D$ channels, we can see positive correlation just beyond the cores. Changes in the cores can also be seen: in the $M_1 \bar{M_1}$ channel, the core becomes larger and softer above $T_c$, soft enough that many pairs can still be found at $x<x_0$, while the cores of the $L$ dyons becomes smaller and harder, reducing the correlation function to 0 at $x<x_0$. The smaller core size also allows the attractive long-range interactions to become larger near to the core, enhancing the correlation found there. 
   
    Despite having no long-range interactions, there is small correlation seen in the same-species ($M_1 M_1$ and $L L$) channels beyond the core due to some mutual correlations to other dyons. Compared the the $D\bar D$ channels the anticorrelation at small $r$ is even stronger due to additional repulsion from the Diakonov determinant. Some correlation functions for the dyons in pure $SU(3)$ Yang-Mills theory at $T/T_c = 0.96$ have been observed in lattice configurations \cite{1908.08709}. We find qualitative agreement with these results in all channels except for dyons of the same species. Here some short-range correlation is observed among the lattice dyons, suggesting an absence of the repulsive core used in our model in this channel.

  \subsection{The vacuum angle and moments of the topological charge}
   
   Another set of important characteristics of the dyon ensemble are its topological properties. Non-abelian gauge configurations generically may possess some topological charge $Q$ with its density $q(x)$ given by 
   \begin{equation}
   q(x)= \frac{g^2}{64 \pi^2} \epsilon_{\mu \nu \rho \sigma}F^a_{\mu \nu}(x) F^a_{\rho \sigma}(x).
   \label{eq_topo}
   \end{equation}
   The (anti)instantons are topologically nontrivial objects with $Q= \pm 1$. The constituent dyons then each carry some fraction of this charge. Dyons of species $i$ carry topological charge $Q = \nu_i$ while antidyons have charge $Q= -\nu_i$. Because we only consider ensembles with equal numbers of dyons and antidyons, $\langle Q \rangle = 0$ in any sub-volume of the box. The most prominent topological feature of the gauge theory, which has been extensively studied on the lattice \cite{Alles:1996nm,Ce:2015qha,Frison:2016vuc}, is the topological susceptibility $\chi = \langle Q^2 \rangle / V_4$. 
   
   Because the dyons in our semiclassical ensemble have well-defined positions and topological charges, measuring the average topological charge only requires summing over the charges of the dyons in a given sub-volume. This is done by splitting the main simulation box in half along each of the 3 axes and computing the total charge in the said half-boxes. This results in 3 independent measurements per configuration, meaning the average of any power of $Q$ is computed from 240,000 measurements. 
   
   \begin{figure}[h]
    \includegraphics[width=0.9\linewidth]{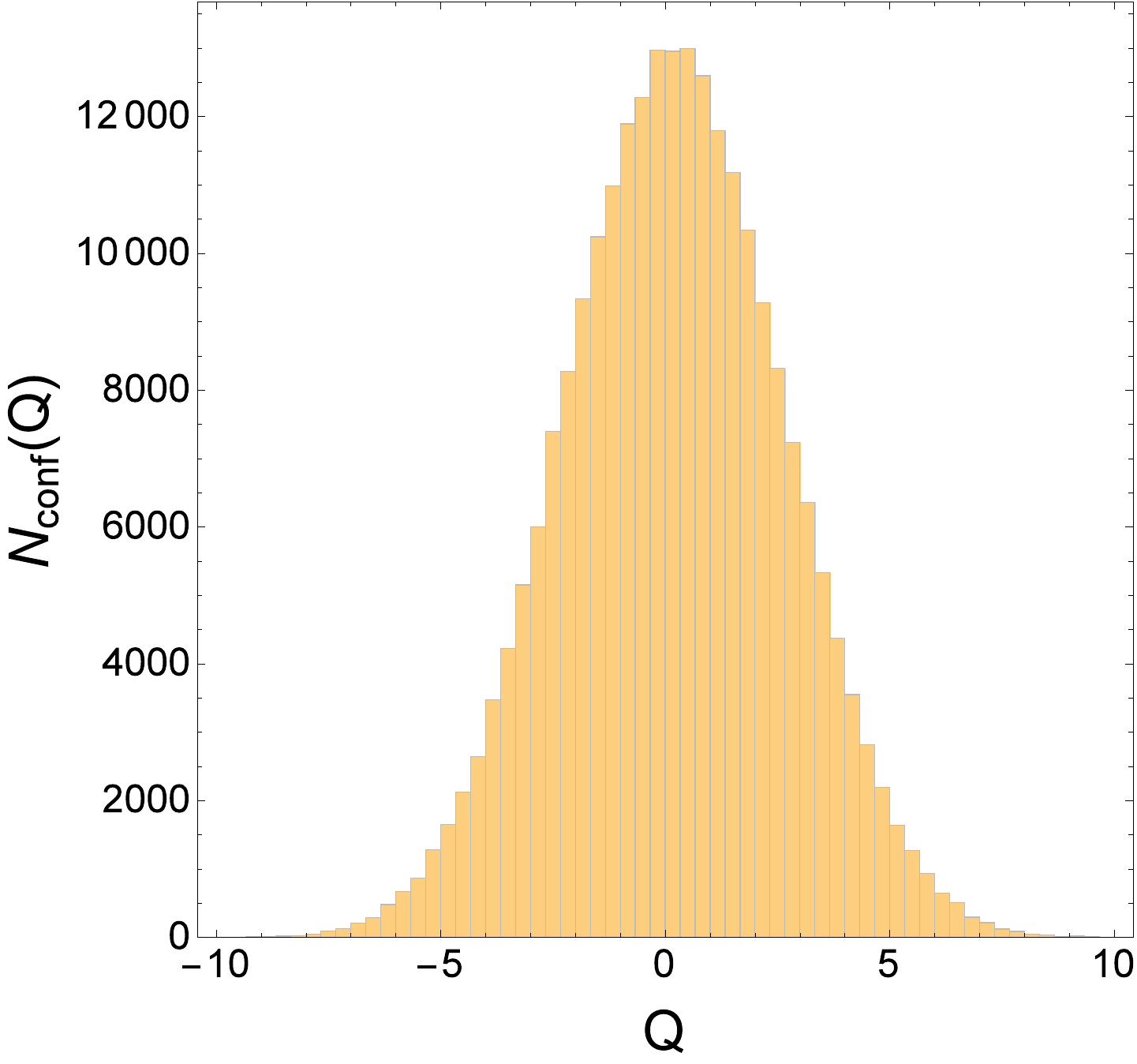}
    \caption{Distribution of values of the total topological charge $Q$ in the half-boxes at $S_0 = 12$ ($T/T_c=0.898$). }
    \label{fig_topo}
   \end{figure}
   
\begin{figure}[h]
      \includegraphics[width=0.9\linewidth]{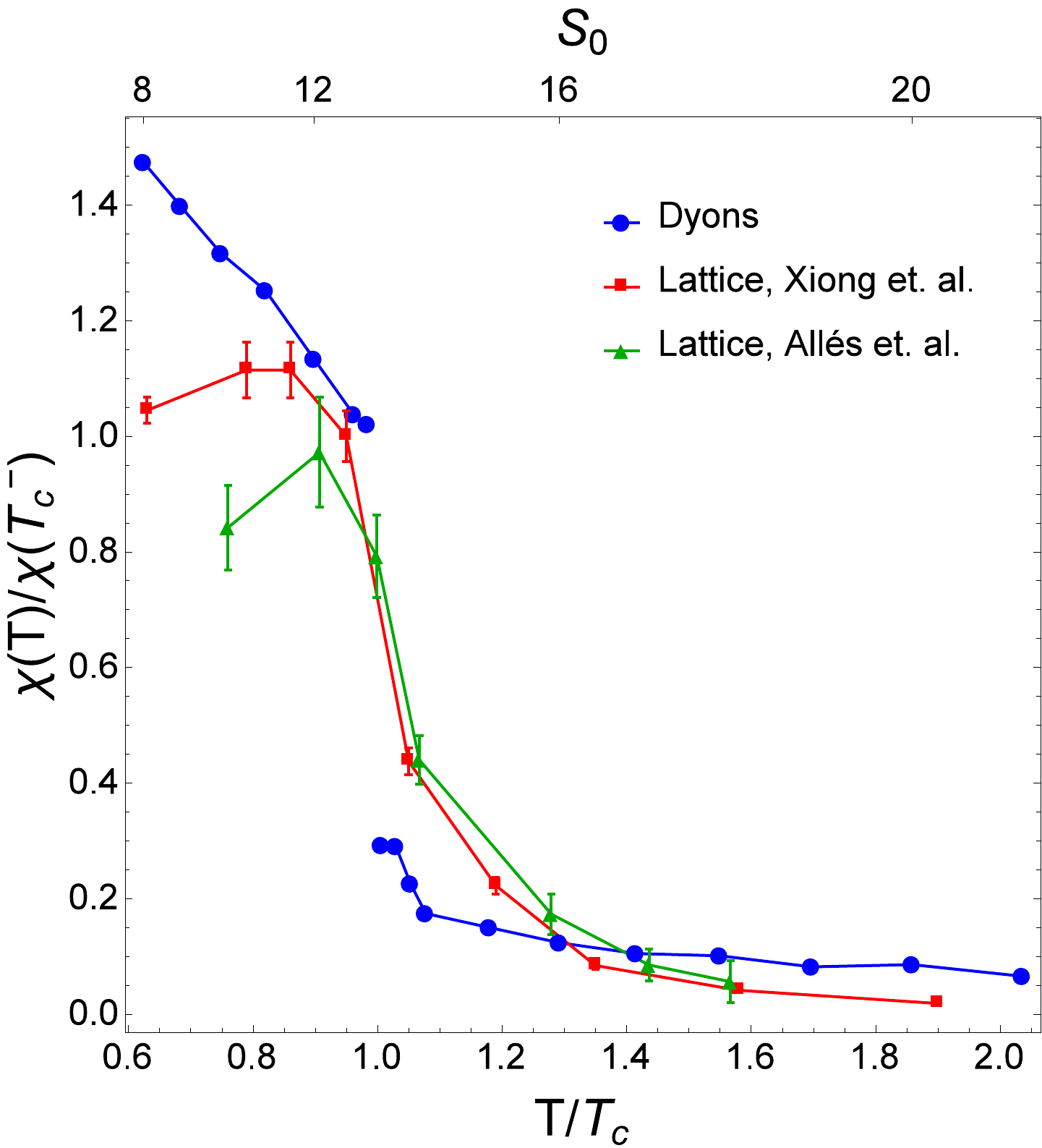}
      \caption{(Color online) Topological susceptibility as function of temperature relative to the topological susceptibility just below $T_c$. Lattice data taken from Refs. \cite{1508.07704, Alles:1996nm}. Lattice values are normalized relative to their stated $T=0$ values. Data from Ref. \cite{Alles:1996nm} (green triangles) is so-called '2-smear' data. Error bars on dyon data are smaller than the point size.}
      \label{fig_chi}
   \end{figure}

   The temperature dependence of the topological susceptibility over the entire temperature range probed in this work is shown by blue points in
   Fig. \ref{fig_chi}. To make comparison to lattice data easier, we show the 
   $relative$ susceptibility, normalized all sets to their values  just below the critical temperature. 
   The issue of absolute comparison is further discussed in Appendix C.
   
   With $\chi (T)$, like $\langle P(T) \rangle$, we see a clear first-order phase transition in the data. Unlike with the Polyakov loop however, such an abrupt transition with a jump is not obvious from the lattice data points (although they of course do not contradict existence of a jump).
   The topological observables have been analyzed on the lattice in many works, we compare our dyon results to those in Refs. \cite{1508.07704,Alles:1996nm}, which cover a similar temperature range as we have studied. 
  
   The (Euclidean) Lagrangians of generic $SU(N)$ gauge theories can be appended by an additional topological term    \begin{equation}
   \mathcal{L} = \frac{1}{4}F_{\mu \nu}^a(x)F_{\mu \nu}^a(x) - i \theta q(x),
   \end{equation}
   where $\theta$ is the so-called vacuum angle and $q(x)$ is the topological charge density as previously defined in Eq. (\ref{eq_topo}). A non-zero $\theta$ explicitly breaks CP symmetry. In QCD it is known that $|\theta_{QCD}| \lessapprox 10^{-10}$, as it is constrained by the upper bound on measurements of the neutron's electric dipole moment \cite{Afach:2015sja,Guo:2015tla}.
   
   By expanding the free energy density $f(\theta)$ around $\theta = 0$, its dependence on $\theta$ can be studied at small, but non-zero $\theta$. It can be expanded as \cite{Vicari:2008jw}
   \begin{equation}
   f(\theta)= f(0) + \frac{1}{2} \chi \theta ^2 (1 + b_2 \theta^2 + b_4 \theta^4 + ...),
   \end{equation}
   where the coefficients $b_n$ are related to cumulants of the topological charge \textit{computed at $\theta = 0$}. We compute the first two terms, which are given explicitly by
   \begin{equation}
   \begin{aligned}
   &b_2 = -\frac{\langle Q^4 \rangle - 3 \langle Q^2 \rangle^2}{12\langle Q^2 \rangle} \\ b_4=& \frac{\langle Q^6 \rangle - 15 \langle Q^2 \rangle \langle Q^4 \rangle + 30 \langle Q^2 \rangle ^3}{360 \langle Q^2 \rangle}.
   \end{aligned}
   \end{equation}
      
   In the confined phase, we find that all $b_2$ values are consistent with $b_2$ being constant below $T_c$. These seven values vary around $~ 0.02 - 0.03$ and have an average $b_2(T<T_c)= 0.026$. Above $T_c$, $b_2$ quickly drops and then remains approximately constant just above $0.01$.
   
   This behavior is in disagreement with available lattice data. The $T=0$ value of $b_2$ has been particularly well studied on the lattice \cite{1109.6815,0705.2352,hep-th/0204125} with values around $b_2(0)=-0.025$. On the high-temperature end, lattice data \cite{1309.6059,1508.07704} finds that $b_2$ approaches the value predicted by the Dilute Instanton Gas Approximation (DIGA), $b_2(T) \rightarrow -1/12$. 
   
   Below $T_c$, our values of $b_2$ are consistent with the magnitude of the $T=0$ value predicted on the lattice, but with opposite sign. In the high temperature limit, we again see that our dyon model predicts positive values rather than negative, as well as being an order of magnitude smaller than the lattice results. Clearly, these small non-Gaussianities in the topological charge distribution are quite sensitive to the dyon interactions. Changes to the dyon interactions could improve the agreement with lattice data. 
      
   \begin{figure}[h]
   	\includegraphics[width=0.9\linewidth]{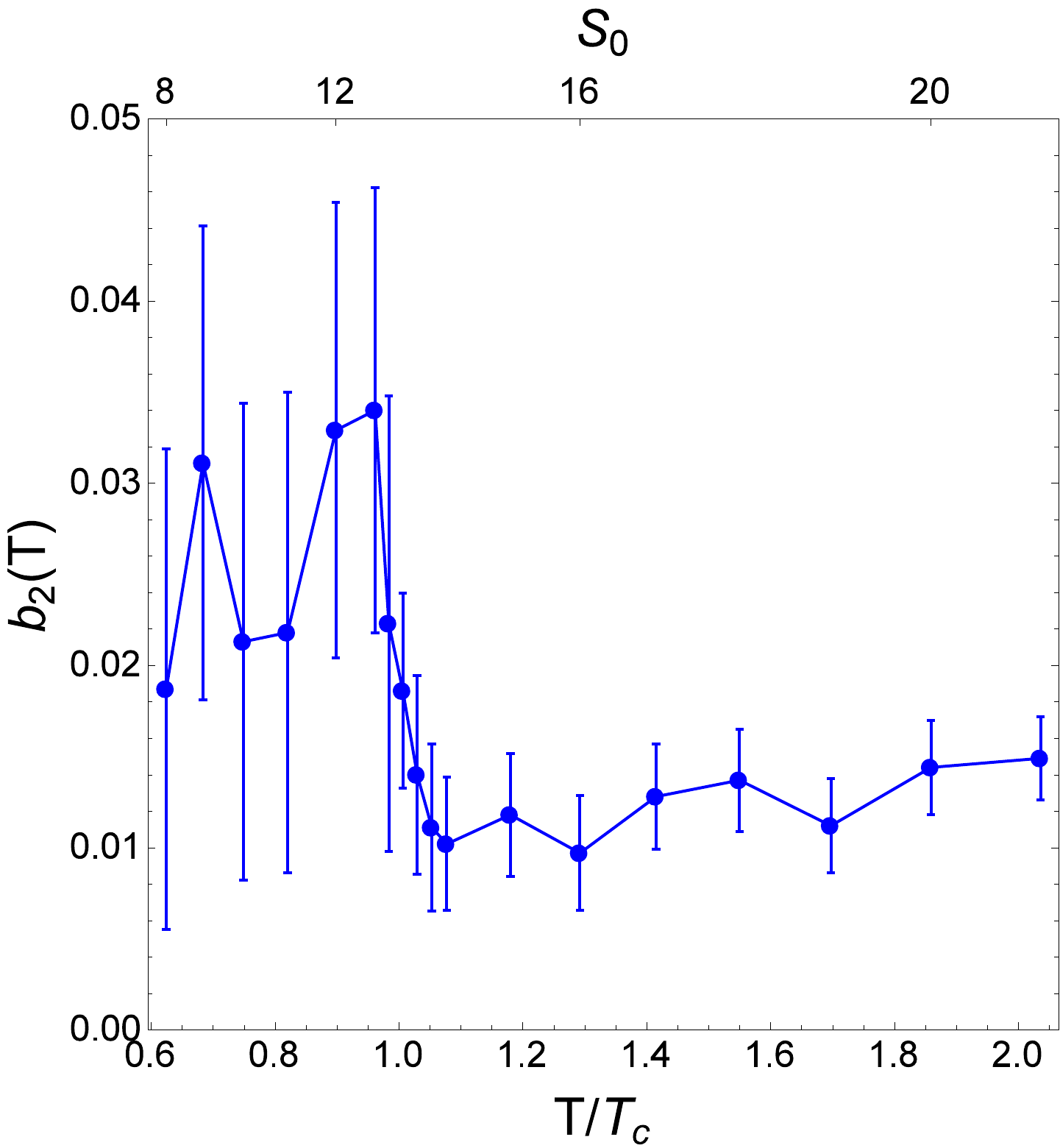}
   	\caption{Expansion coefficient $b_2$ as a function of temperature.}
   	\label{fig_b2}
   \end{figure}    
   
   Finally, our values for $b_4$ are compatible with zero for all temperatures as is also observed on the lattice \cite{1109.6815,1512.01544}. Upper limits from these lattice results constrain its value to be $|b_4(T=0)| < 10^{-3}$.   
   
   \section{Dyons in the trace-deformed gauge theory}
   
   \subsection{Polyakov loop and the phase transition}
   Trace-deformed gauge theories (TDGT) were introduced 
   in Refs. \cite{Myers:2007vc,Unsal:2008ch}, adding certain terms containing powers of the Polyakov line to the theory action, such as 
   \begin{equation}
   \Delta S_{def}= h \int d^3x | P(\vec x) |^2
   \end{equation}
   with a new parameter $h$. At $T>T_c$, when in the un-deformed
   theory the Polyakov loop is nonzero, the new term
   obtains an additional contribution and suppresses 
   its value, shifting the theory back toward the confining holonomy. For large enough $h$ the system returns to
   the $\langle P \rangle =0$ phase, which we will call
   the ``reconfined phase".
   
   These theories  have been studied in numerous lattice simulations, of which we will
   mention those of the Pisa group \cite{Bonati:2018rfg,Bonati:2019unv,Athenodorou:2020clr}. Their main findings are that for a specific temperature above $T_c$ at large enough $h>h_*\sim 1/3$, at which  $\langle P \rangle$ returns to zero, there is no more dependence on $h$, and the reconfined phase is remarkably similar to
   the original (low temperature) confined phase. It was
   shown to
   possess the same spectrum, topological observables, and even spectrum of periodic strings (torelons).  
   
   From the perspective of the instanton-dyon theory that
   we are developing, it is clear that the reconfined phase,
   like the confined one, corresponds to symmetric value of the holonomy $\nu=1/3$ and the same densities
   of all species $L,M_1,M_2$ of the dyons. Yet the
   temperature scale, and thus the periodicity
   interval of the Euclidean time $\tau$, is different.
   Therefore, the absolute scale of the dyon action
   parameter $S_0=8\pi^2/g^2(T)$ must increase, making
   the system more dilute.  And yet, the topological susceptibility $\chi$ is the same as at $T=0$ despite a significantly reduced dyon density! 
   
   It is possible to study the instanton-dyon ensemble in a theory with arbitrary $h$ by the addition of a deformation term to the free energy density $\Delta f_{def} = h \langle P \rangle ^2$. Because this new term depends only on the value of the holonomy, one does not need to perform new MC simulations to study the effect of changing $h$; one needs only to add the new term to the un-deformed results and perform new fits to find the minima.
   
\begin{figure}[h]
      	\includegraphics[width=0.9\linewidth]{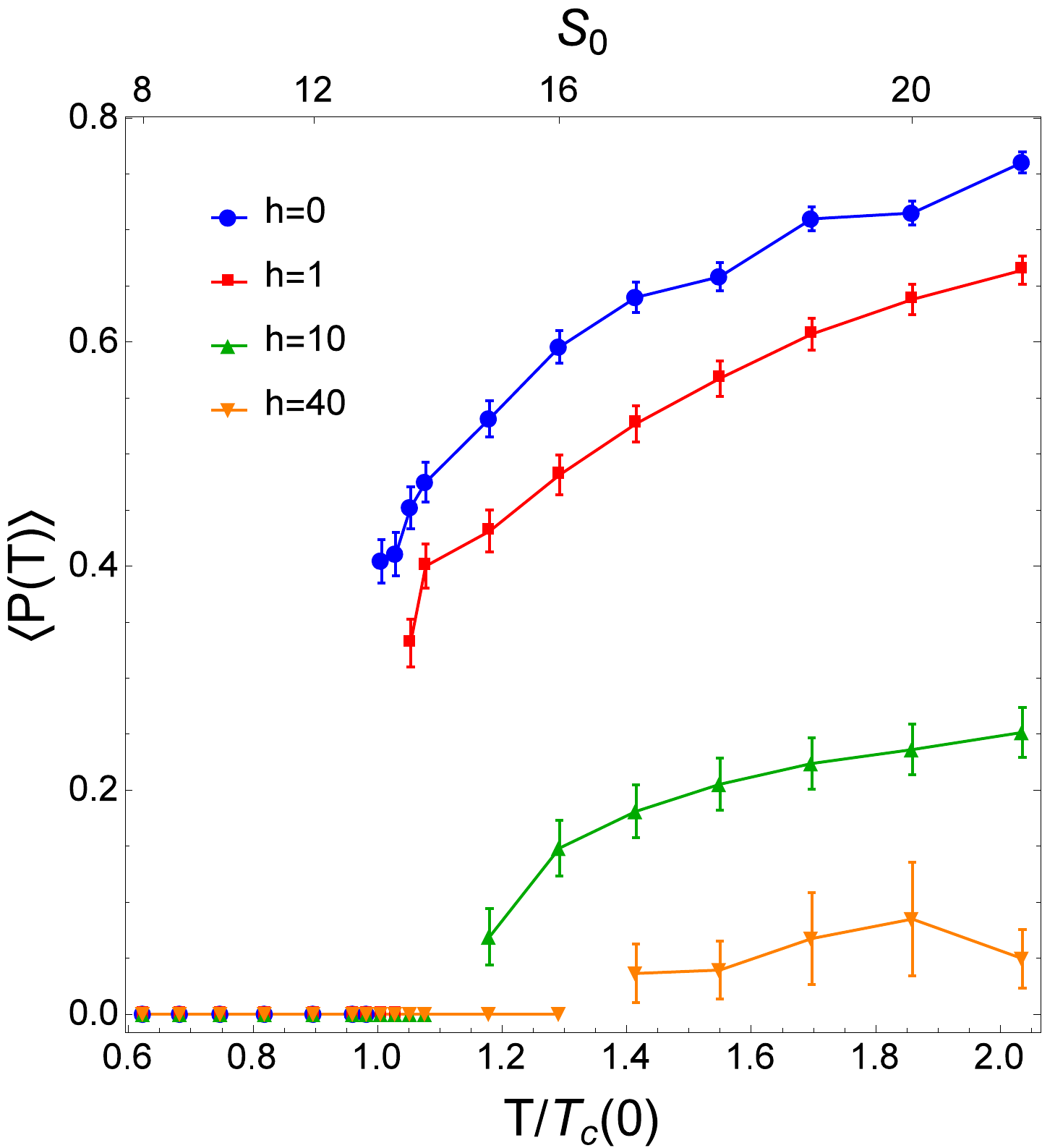}
      	\caption{(Color online) The value of the Polyakov loop as a function of temperature $\langle P(T) \rangle$ for different values of the trace-deformation parameter $h$. }
      	\label{fig_polyh}
      \end{figure}

   The first questions to ask are how does changing $h$ affect the value of the Polyakov loop and at what value of $h$ does the system become reconfined? Fig. \ref{fig_polyh} shows the temperature-dependence of the Polyakov loop for a few different values of $h$. As expected, the larger the trace deformation, the more suppressed the values of $\langle P \rangle$ become at $T>T_c$. Additionally, we see that increasing $h$ increases the critical temperature of the theory as well, as the Polyakov loop is suppressed back to the confining value. 
   
   The critical temperature of the trace-deformed theories are determined via the same method as the original theory: the intersection of linear fits to $f$ on both sides pf the transition. At this point we are limited to determining $T_c$ for $h \le 40$, as above this value the the Polyakov loop remains very close to the confining value and the minimum value of the holonomy fit become closer to $\nu = \frac{1}{3}$ than the holonomy step size used for the simulations, meaning that $\nu= \frac{1}{3}$ is compatible with our uncertainties for all temperatures studied. In order to accurately determine the critical temperature for large $h$, smaller steps in both $S_0$ and $\nu$ are necessary. This also makes determining the nature of the phase transition inconclusive at larger $h$. At small $h$, it is clear to see that the phase transition remains first order, while at larger $h$ (above $\sim 10$) we do not have sufficient resolution in $\nu$ to determine whether the holonomy potential continues to have two distinct minima or becomes a smooth crossover.
   
      \begin{figure}[h]
         \includegraphics[width=0.9\linewidth]{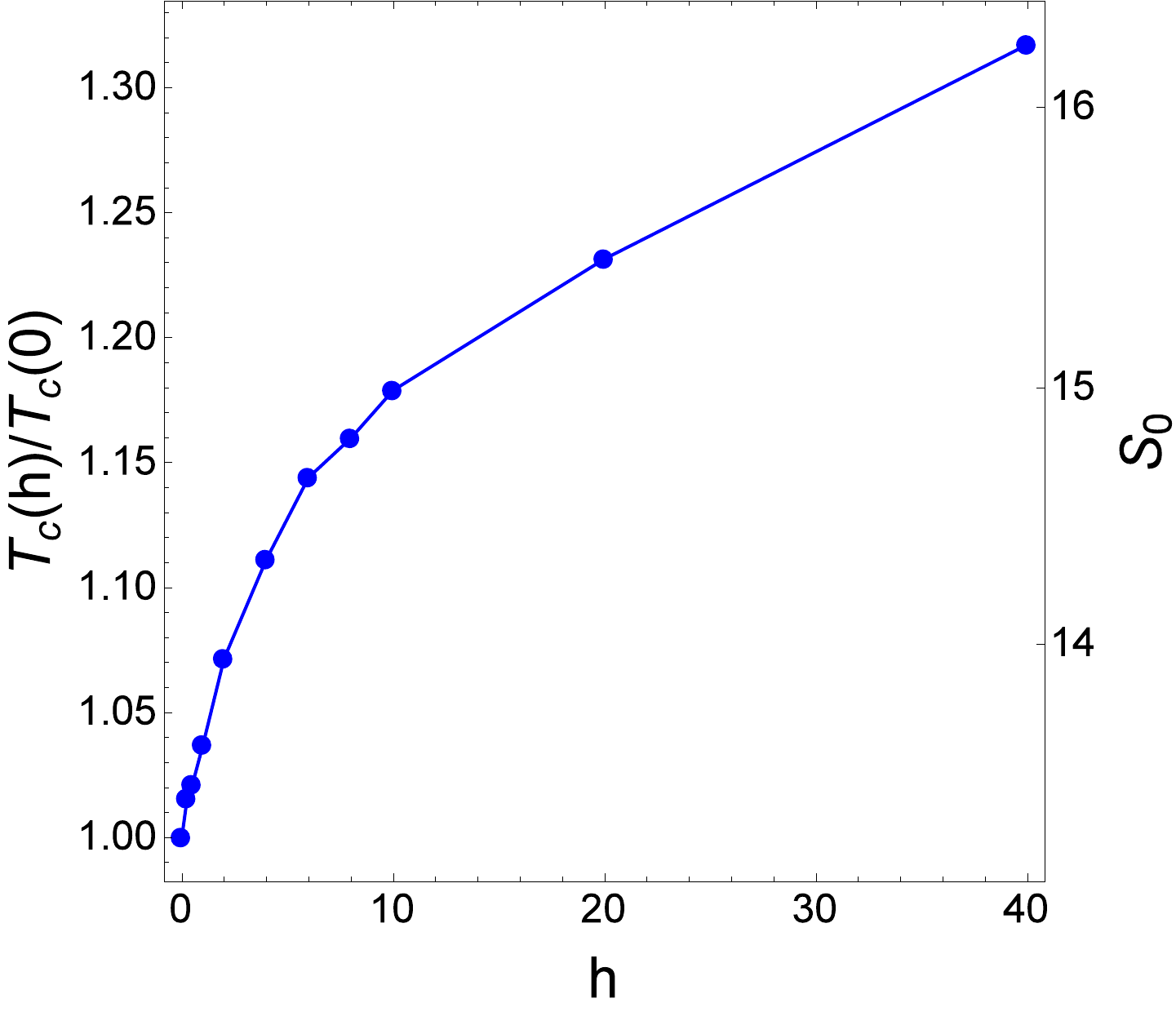}
         \caption{Critical temperature of the theory with trace-deformation parameter $h$ relative to that of the the pure Yang-Mills theory.}
         \label{fig_crith}
      \end{figure}
   
   While we are currently limited in how high of a critical temperature we can probe (up to $\sim 1.3T_c(0)$ as per Fig. \ref{fig_crith}), the theory may have non-trivial behavior at higher temperatures. In the limit that $T \rightarrow \infty$ the density of dyons goes to zero and only two terms in the free energy density remain: the (deconfinement-favoring) Gross-Pisarski-Yaffe potential and the (confinement-favoring) trace-deformation term. Clearly which term dominates depends on the value of $h$. At $h > 5 \pi^2/18 \simeq 2.74$ the confining holonomy is the global minimum. 
   
   This leads to the conclusion that there are three distinct regimes of the TDGT: \\i) $h <5 \pi^2/18$: The theory is confined at low temperatures and deconfined at high temperatures. \\ii) $ 5 \pi^2/18 < h < h_{max}$: The theory is confined at low temperatures, deconfined in some intermediate region and then \textit{confines again} at high temperatures. \\iii) $h > h_{max}$: The theory is confined at all temperatures.
   
   While we can't yet access high enough temperatures to directly see the second phase transition in regime (ii), the last value of $\langle P \rangle$ for $h=40$ in Fig. \ref{fig_polyh} suggests that $\langle P \rangle$ may be starting to decrease and return to zero. 
   
   \subsection{Topological observables of the reconfined theory}
   
   One of the most interesting features of the deformed gauge theory observed on the lattice \cite{Bonati:2018rfg,Bonati:2019unv,Athenodorou:2020clr} is that when the trace deformation is large enough to return the system to the confining holonomy, the topological observables, namely $\chi$ and $b_2$, also return to the same values they had in the confined phase of the un-deformed theory, and then show no more dependence on further increasing $h$. 
   
   Rather than choose specific temperatures and vary $h$ as was done in the lattice studies, we choose an arbitrarily-large value of $h$ and vary temperature. As mentioned in the previous section, for some large value of $h$, such that $h>h_{max}$, the system should remain in the confined phase at all temperatures. In our dyon model, we can study this scenario, which we call the 'maximally-deformed' theory, by simply demanding that $\nu = \frac{1}{3}$ and determining the dyon densities that minimize the free energy density with this constraint. Because our individual simulations have fixed holonomy value $\nu = \frac{1}{3}$, $\langle P \rangle = 0$ the trace-deformation term does not contribute to the free energy density of these simulations and doesn't need to be considered in the fits. 
 
   \begin{figure}[h]
   	\includegraphics[width=0.9\linewidth]{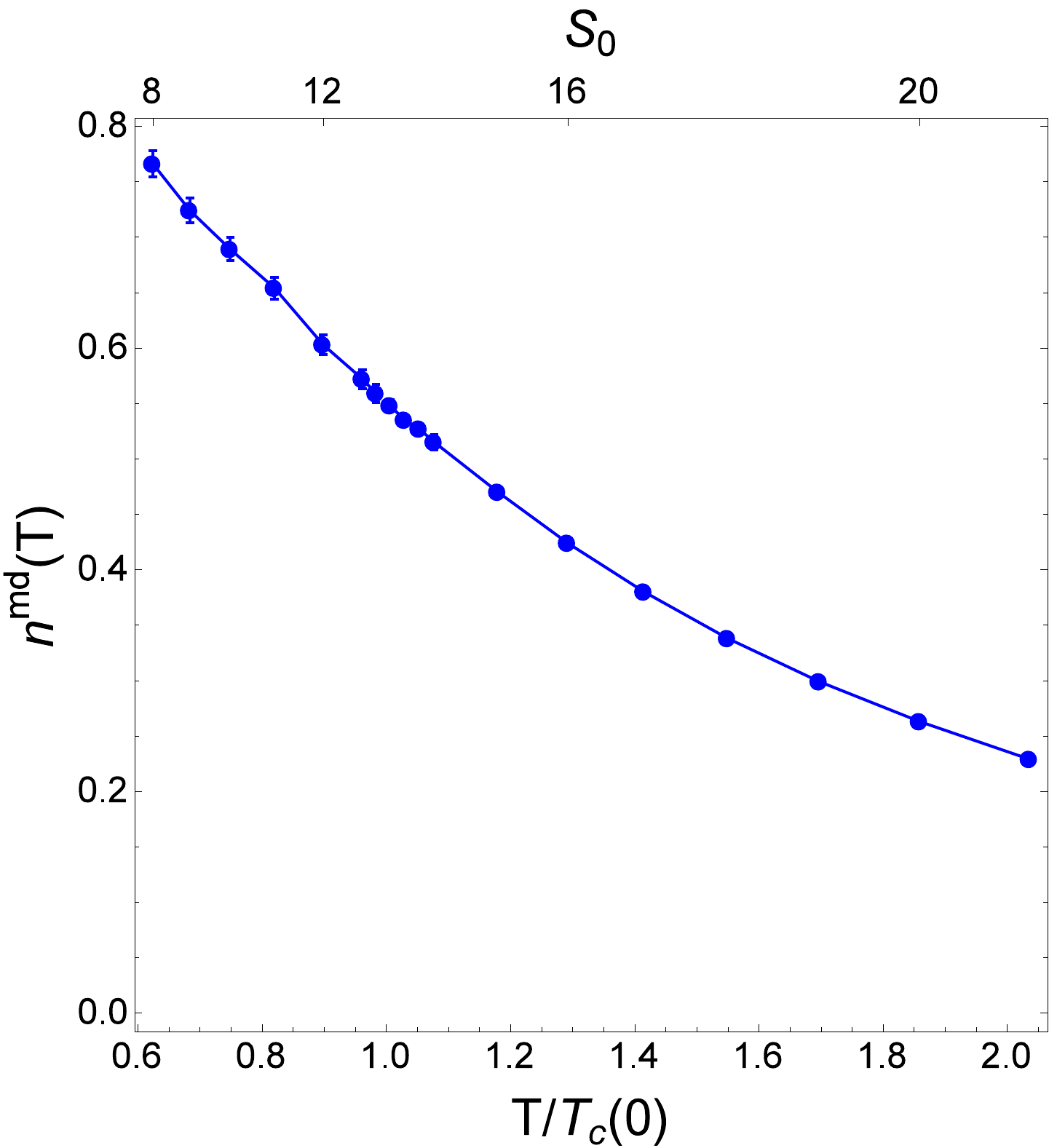}
   	\caption{Dyon density of the maximally-deformed theory as a function of temperature. Both dyon densities are equal $n = n_M = n_L$.}
   	\label{fig_ddef}
   \end{figure}

   \begin{figure}[h]
      \includegraphics[width=0.9\linewidth]{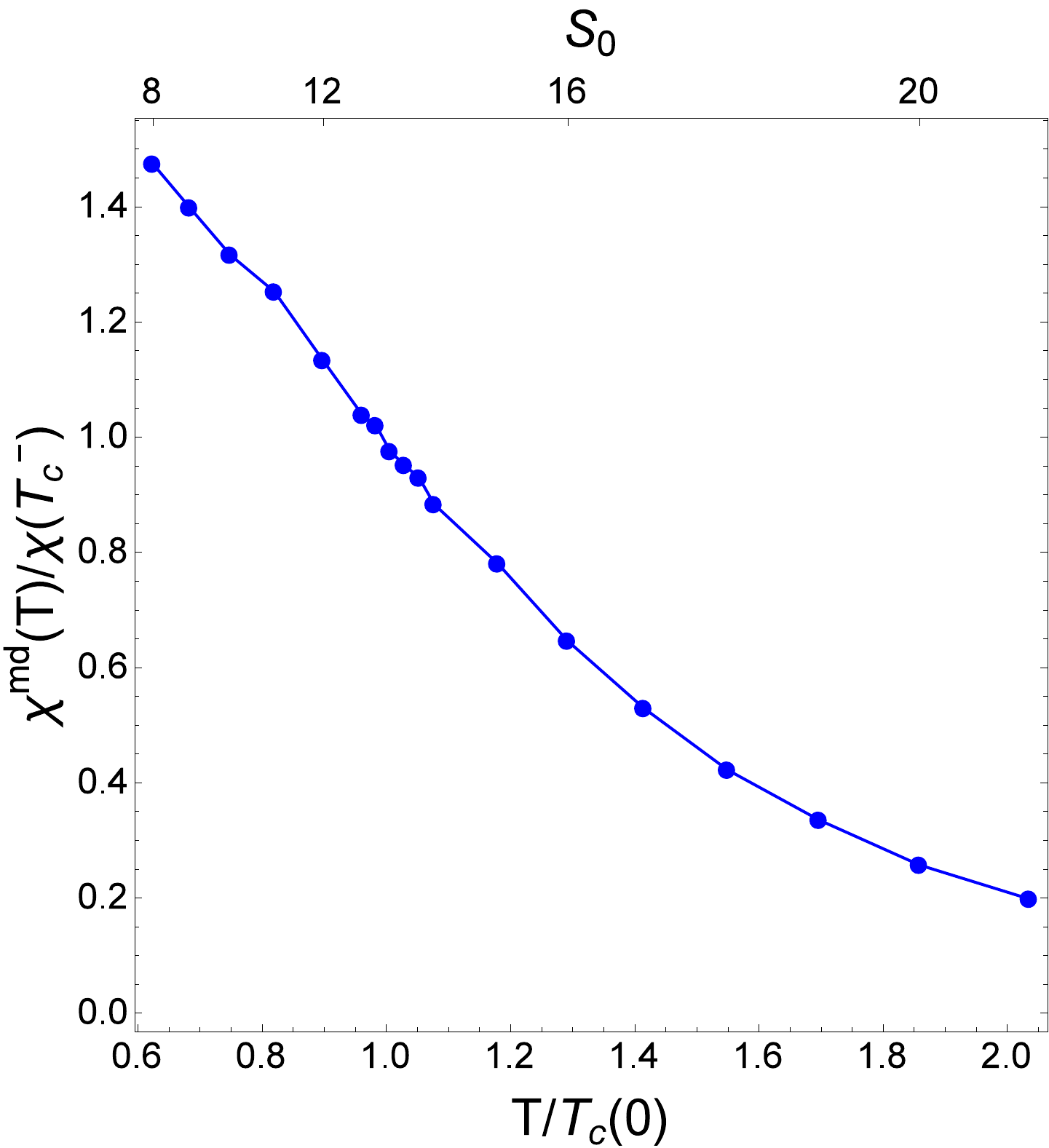}
      \caption{Topological susceptibility of the maximally-deformed theory as a function of temperature. Error bars are smaller than the point size.}
      \label{fig_chidef}
   \end{figure}
   
   The question then is whether or not the dyon ensemble, like the lattice theory, possesses the same topological observables in the maximally-deformed theory as in the original confined phase. In the case of $\chi$, our dyon model did not predict a constant value but an approximately linearly-decreasing one. We observe analogous behavior to what is seen on the lattice; in the maximally-deformed theory the abrupt jump in $\chi$ disappears and it displays the same temperature dependence above $T_c(0)$ as below. Comparing the dyon density (Fig. \ref{fig_ddef}) and the topological susceptibility (Fig. \ref{fig_chidef}), one can see that the decrease in $\chi$ is mostly driven by the decreasing dyon density. Although $n_i$ decreases by a factor of $\sim 4$ while $\chi$ decreases by a factor of $\sim 7$ over the temperature range we study, meaning the value of $\langle Q^2 \rangle$ decreases by a factor of $\sim 2$ as we indeed observe. 
   
   The behavior of $b_2$ is more inconclusive. It remains constant up to about $1.1T_c(0)$ and then decreases quickly and is compatible with zero at all temperatures above that point. Again, it remains positive at all temperatures. It should be noted that the error bars on the $b_2$ measurements are still quite large and the trend could look quite different with improved statistics.
      
     \begin{figure}[h]
        \includegraphics[width=0.9\linewidth]{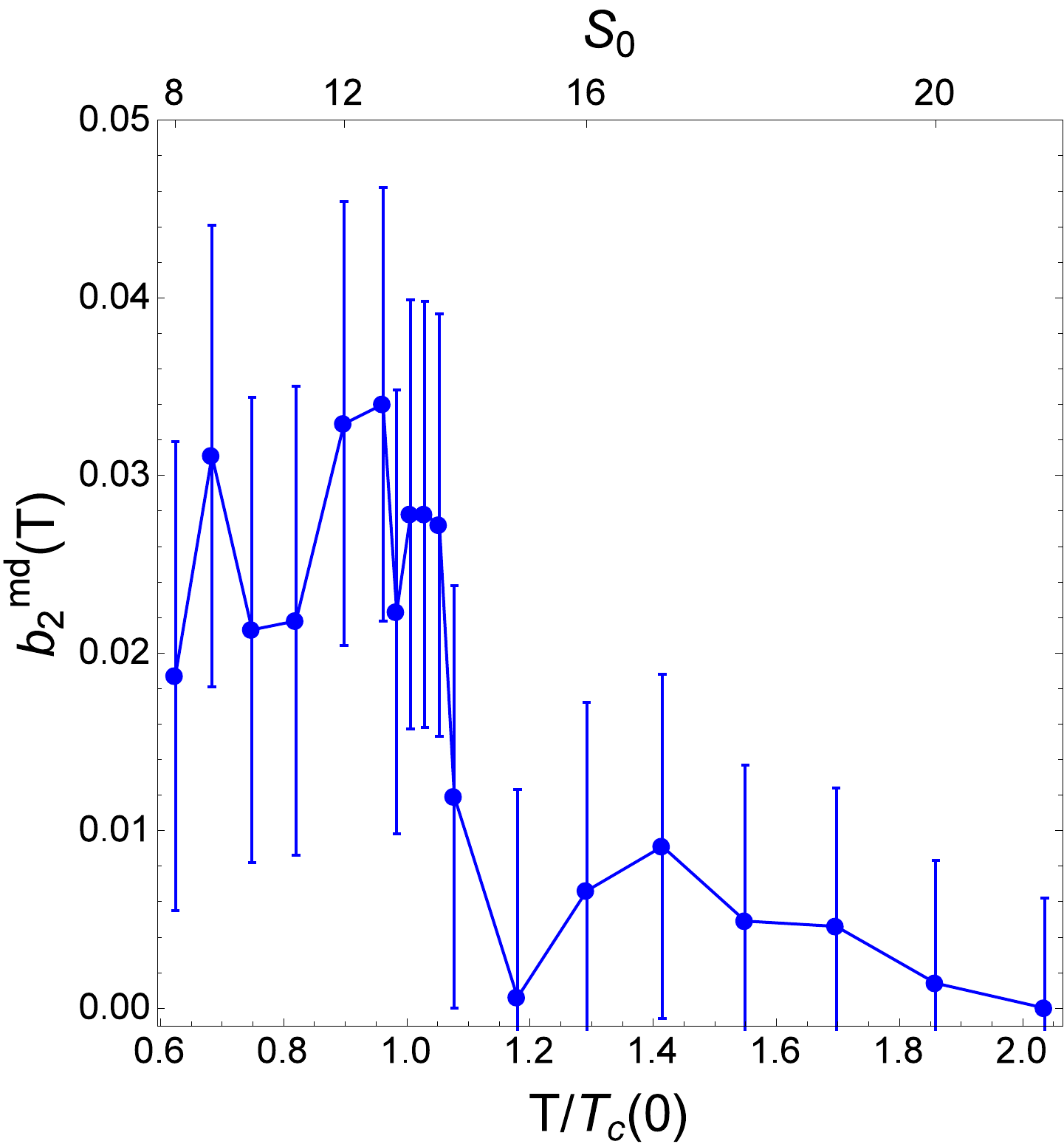}
        \caption{Expansion parameter $b_2$ in the maximally-deformed theory as a function of temperature.}
        \label{fig_b2def}
     \end{figure}
   
\begin{figure}[t]
        \includegraphics[width=0.9\linewidth]{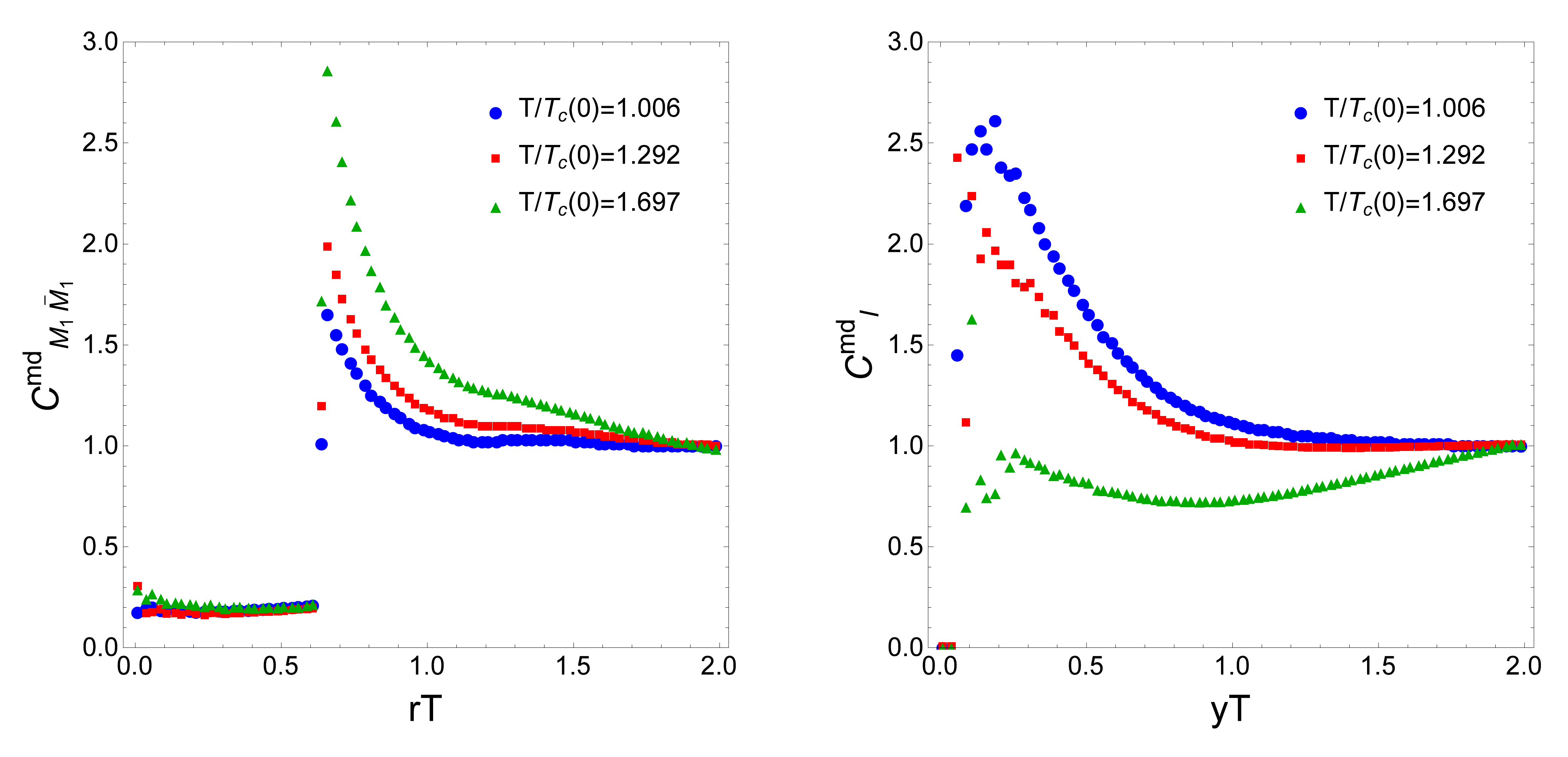}
        \caption{(Color online) Spatial correlation functions of the dyons in the $M_1 \bar{M}_1$ (left) and instanton (right) channels in the maximally-deformed theory at three different temperatures. }
        \label{fig_corrdef}
     \end{figure}

   In order to determine what drives this behavior, it is useful to look at some correlation functions in this deformed theory. It is clear that the ensemble has some non-trivial temperature dependence even though it remains in the confined phase. The two main channels with positive correlation are, just like the un-deformed theory, the $D\bar D$ channels and the instanton  ($M_1M_2L$) channel. As the temperature is increased, the ensemble more strongly prefers $D \bar D$ pairs rather than (anti)instanton as can be seen in Fig. \ref{fig_corrdef}. This is due to the increase in the strength of the $D\bar D$ classical attraction. Eventually, these pair correlations grow strong enough to completely kill correlation in the instanton channel.  
   
   In the original theory, the system prefers binding into instantons
    in the confined phase and then transitions into preferring $M_1 \bar M_1$- and $M_2 \bar M_2$ pairs at high temperatures; the density of $L\bar L$ is becoming small at this point. However, because the core sizes of the $M_i$ dyons grow large, the binding in these channels does not become too large.
   
   The maximally-deformed theory then displays behavior not seen in the original theory. At high temperature the system becomes a three-component  gas of $D\bar D$ pairs, with all  types having equal densities and core sizes. Because the theory remains confined, the core sizes never become large and the correlation between dyons in $D\bar D$ pairs becomes very large. The topological observables are very sensitive to this binding , as the (anti)instantons and $D\bar D$ pairs have topological charges $Q=\pm 1$ and $Q=0$, respectively. The shift towards the latter leads to a \textit{neutralization} of the topological charge and is responsible for driving the values of $\chi$ and $b_2$ down at higher temperatures.   
   
   Looking to the future, one can see that this behavior has interesting implications when quarks are added to the theory. It is known that chiral symmetry restoration is directly related to the quark zero modes on the instantons. In particular, the quark interactions drive the system to form instanton-antiinstanton 'molecules'. This neutralizes the fluctuations of the topological charge and
  shifts the Dirac eigenvalues away from zero 
at high temperatures, restoring chiral symmetry \cite{Ilgenfritz:1994nt}. 
In fact in this work we already observe   $D\bar D$ pairing at higher temperatures. 
We see here that such correlations continue to grow, even when the theory remains confined due to the
deformation term in the action.
 This suggests that in the trace-deformed theory with quarks one may see chiral symmetry restoration occur $inside$ the confined phase.       
         
    \newpage     
    \onecolumngrid   
         
     \twocolumngrid     
   \section{Summary and discussion}

   This paper reports the first direct numerical simulations of the ensemble of
   instanton-dyons in the pure $SU(3)$ Yang-Mills theory. We use classical and semiclassical one-loop
   measures which have derived the inter-dyon interactions. We have performed Monte-Carlo simulations of an ensemble of dyons in a 3D cube with periodic boundary conditions over a range of temperatures, densities, and holonomy values.
   
   Our main objective is to see whether the semiclassical ensemble of instanton-dyons does or does not
   reproduce the deconfinement transition, which is, in this case, of the first order. We indeed find that
   there are two distinct phases, the confined (with free energy minimal at holonomy value $1/3$,
   Fig. \ref{fig_struc} (left)) and the deconfined (with free energy minimal elsewhere, Fig. \ref{fig_struc} (right)). We were able to map out the holonomy potential and see explicitly the two nearly-degenerate minima of the potential representing the transition between the two phases (red squares in Fig. \ref{fig_struct2}) and study various properties near the critical temperature.
   
   Comparison between our semiclassical model and lattice data for VEV of the Polyakov loop $\langle P(T) \rangle$ is
   shown in Fig. \ref{fig_poly}. Remarkably, the jump of it at $T_c$ is reproduced rather precisely, with some
   moderate deviations at higher temperatures. As shown in Fig. \ref{fig_dens}, this large jump results in
   $L$-dyon suppression in the deconfined phase, while the density of $M_i$-dyons has a rather smooth temperature dependence.
   
   The pair-wise spatial correlations between various dyons are shown in Fig. \ref{fig_corr}. Overall, they are in qualitative
   agreement with expectations. Note also, that for a system with a strong first order transition, one does not see drastic changes in the correlations, except for overall change of scale due to the change
   in the temperature. 
   
   We have studied fluctuations of the topological charge. Since our simulations each have a fixed
   number of dyons, this is done by counting charges in the half-box. A typical distribution is shown in Fig. \ref{fig_topo}, at first sight having just a normal Gaussian form.
   Its width - the topological susceptibility $\chi$ - is compared to lattice data in Fig. \ref{fig_chi}. The magnitude of the jump and behavior in the deconfinement phase are found to be quite similar, but the dependence at $T<T_c$ is somewhat different. Our results for non-Gaussianity parameter $b_2$ are perhaps still too noisy. Let us note that these observables, as well as the dyon correlations, are very sensitive to the details of the dyon interactions, particularly the short-range classical interactions, which are the strongest. These interactions are at present the most poorly-understood aspect of the dyon model with some parts being analytically derived (e.g. the Diakonov determinant) and others being simply phenomenological (e.g. the repulsive core). In order to achieve a quantitative agreement between the dyons and lattice data, if possible, these interactions should be the subject of further study and improvement. 
   
   We also report the first study of trace-deformed gauge theory in the instanton-dyon 
   setting. As expected, by increasing the coupling $h$ of the deformation term,
   one indeed suppresses the jump and value of the Polyakov loop in the deconfined phase.
   The phase transition location is also pushed to higher values, see Fig. \ref{fig_crith}. Overall,
   like on the lattice, we see that increasing $h$ leads to a return to the topological susceptibility as in the confining phase of the un-deformed theory.  We however still see that this agreement is only partial, see for example the spatial correlations taken in the maximally-deformed confining theory.
   
   Completing this summary of our results, we again emphasize that the semiclassical model
   used, which has many orders of magnitude fewer degrees of freedom than lattice gauge theory, is able to explain properties of the deconfinement phase transition
   of the $SU(3)$ gauge theory quite well. Some quantities -- like the jump in the Polyakov line -- are reproduced precisely, others semi-qualitatively, but we have not seen any
   serious disagreements. The double-trace deformation study also shows that
   both our model and lattice react to this addition to the action in a very similar way.
   
   Thinking of the future, the next step will of course be the inclusion of dynamical quark flavors via their zero modes and interactions. Such work will allow not only for the study of confinement (which will now be expected to be a smooth crossover), but also to study the role of the dyons in chiral symmetry restoration. 

   \begin{acknowledgments}
	This work is supported by the Office of Science, U.S. Department of Energy under Contract No. DE-FG-88ER40388. The authors also thank the Stony Brook Institute for Advanced Computational Science for providing computer time on the SeaWulf computing cluster. 
   \end{acknowledgments}
   
    \appendix
    \section{Units}
    Temperature is the main physical quantity of this work: it defines the holonomy via $A^3_4 = 2 \pi \nu T$ as well as the physical sizes of the dyons. As seen in previous equations, all interactions of the dyons contain the dimensionless distance $rT$ rather than the dimensionful distance $r$. Because of this we define many quantities in terms of this dimensionless distance: the dimensionless 3-volume $\tilde{V}_3 = V_3 T^3$, the dyon densities $n_i = \frac{N_i}{\tilde{V}_3}$, and the free energy density $f = \frac{F}{\tilde{V}_3 T}$. 
    
    All parameters are presented dependent on the instanton action $S_0$ which is related to the temperature by
    \begin{equation}
     S_0 = \frac{8 \pi ^2}{g^2} = (\frac{11}{3}N_c - \frac{2}{3}N_f) \ln \left( \frac{T}{\Lambda} \right)
     \label{eqtemp}
     \end{equation}
    where, in pure $SU(3)$ gauge theory, $N_c=3$ and $N_f=0$.
    
    As with previous work on the ensemble of $SU(2)$ dyons \cite{Larsen:2015vaa}, the value of the parameter $\Lambda$ is a choice that maps the instanton action $S_0$ to the temperature $T$. This parameter was chosen to be $\Lambda = 2.8$. It should be the subject of improvement in future comparison with lattice data. The reader should keep in mind that this value is different than the one chosen for the previous $SU(2)$ studies, $\Lambda^{SU(2)}=1.5$. Because the temperature scale is set by this \textit{choice}, conversion to physical units may be done setting the dyons' critical temperature equal to that of the lattice gauge theory: $T_c =9.82 \leftrightarrow 260$ MeV \cite{Boyd:1996bx,Borsanyi:2012ve}. 
    
    \section{Comments on interaction parameters and finite-size effects}
    
    \subsection{Parameters of the dyon interactions}
    Let us take a moment to remind the reader that this is a \textit{model} of the dyon ensemble. Certain parameters of the dyon interaction, namely $V_0$, $x_0$, and $M_D$, are not known from first principles, and are thus phenomenological parameters of the model. In the previous $SU(2)$ work \cite{Lopez-Ruiz:2016bjl} these parameters were chosen to be $(V_0,x_0,M_D) = (20,2.0,2.0)$. In preliminary of testing of the $SU(3)$ model, it was found that these parameters were unsuitable for generating the desired phase transition. These parameters are sensitive to the details of the theory and one should not expect that they are identical for all $N_c$. 
    
    In this work, the parameters used were $(V_0,x_0,M_D) = (10,4/3,1.5)$. These parameters were chosen to produce a phase transition around $S_0 \sim 12$, jumping from $\langle P \rangle = 0$ to $\langle P \rangle \simeq 0.4$. The choice of $x_0$ was not arbitrary, as when the factor of $1/\nu$ is included, the dyons in the confined phase have the same dimensionless core radius as was used in $SU(2)$. This choice of parameters exists on a 3D 'island' in the parameters space in which such choices of the parameters possess reasonable properties. They can and should be further constrained by future data regarding identification of the dyons in lattice configurations. In particular, if dyon densities and correlations can be studied with sufficient statistics, such data can reveal details of the dyons' interactions. Some correlations between dyons on the lattice (in physical QCD) can be found in Ref. \cite{Larsen:2019sdi}, although the statistics are not yet sufficient to make such constraints. 
    
    Each of these parameters has a clear effect on the phase transition of the ensemble. Large values of any of these parameters favors the confined phase and drives $T_c$ upward. In the case of $V_0$ and $x_0$, large values make the repulsive cores larger, more strongly favoring the confining holonomy. A large value of $M_D$ suppresses the large-range interactions making the dyon cores comparatively more important. For a more systematic analysis of the effects of changing the parameters see Ref. \cite{Lopez-Ruiz:2016bjl}.
    
    Additionally, the Debye mass is not necessarily a constant, but rather a temperature-dependent quantity $M_D(T)$. In fact, in the previous work by one of the authors \cite{Larsen:2015vaa} the Debye mass was treated as an input parameter to be swept over, much like the densities and only configurations with input values consistent with the its definition
    \begin{equation}
    M_D^2 =  \frac{g^2}{2V} \frac{\partial^2 F}{\partial \nu ^2} |_n,
    \end{equation}  
    were chosen. This increases the number of configurations to be tested, but allows for a more general, temperature-dependent Debye mass. Both methods have been shown to produce reasonable properties for the $SU(2)$ dyon ensemble, so we have saved computational time by treating it solely as an input. One could also consider a more 'direct' method in the future, simply taking the value of $M_D(T)$ as an input from lattice data. 
    
    \subsection{Finite-size effects}
    
    In principle, equilibrium statistical mechanics is derived in the thermodynamic limit $N_D, \tilde{V}_3 \rightarrow \infty$, which is not possible in numerical simulations. For such a finite numerical simulation, the main concern is whether the system size used is sufficiently large enough so that the measured observables can be reliably extrapolated to the thermodynamical limit. The simplest way to check these effects is to simulate larger systems with the same parameters and study the dependence of some observables on the size $N_D$. 
    
\begin{figure}
       	\includegraphics[width=\linewidth]{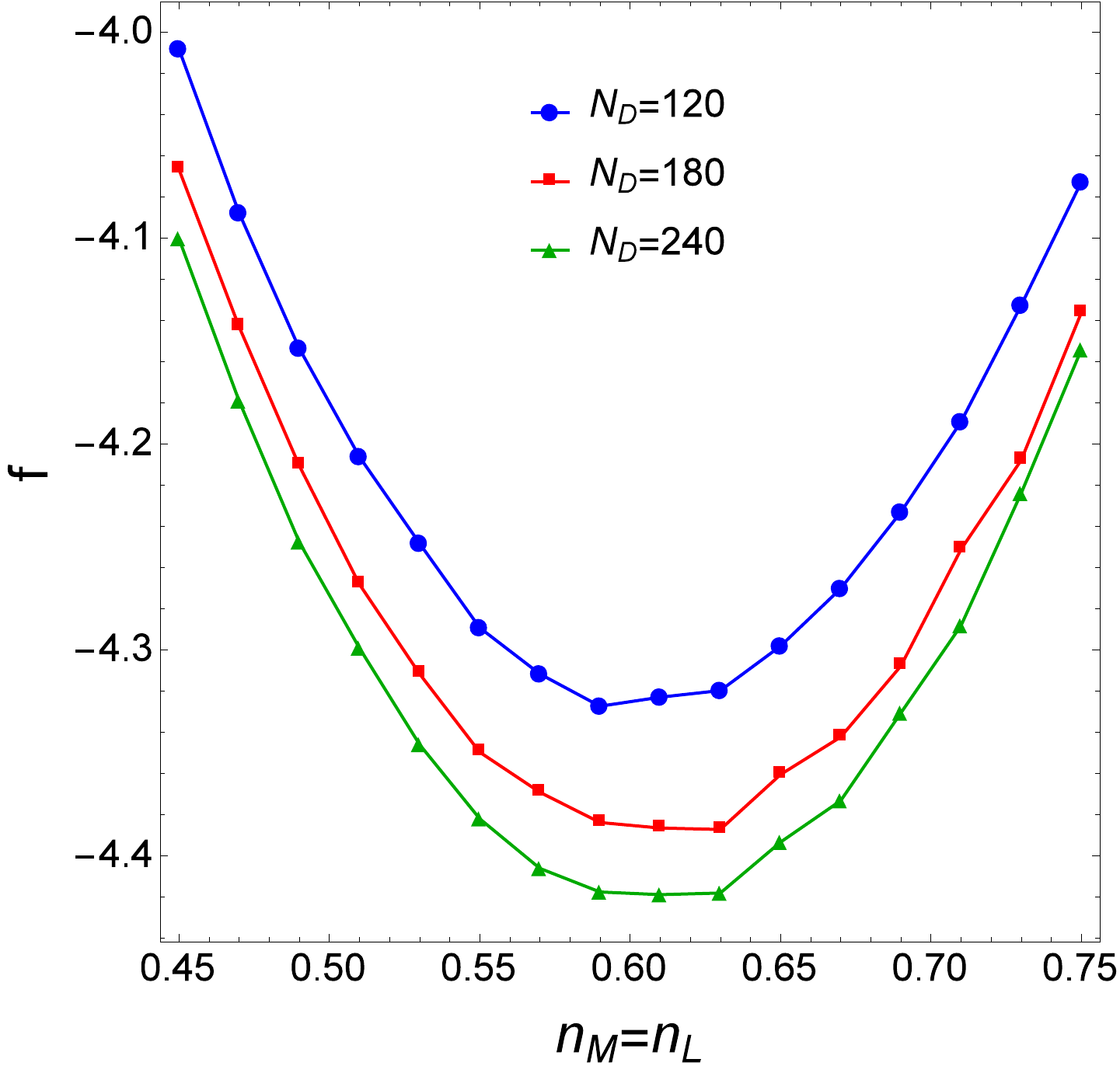}
       	\caption{(Color online) Free energy density $f$ as a function of dyon density with $S_0 =12$, $\nu= \frac{1}{3}$ for three different system sizes.
       	}
       	\label{fig:size}
       \end{figure}

    Here we discuss a representative example of the finite-size effects as seen in Fig. \ref{fig:size}. In increasing the size of the system by up to a factor of 2, there is a noticeable change in the free energy landscape. Looking at Eq. (\ref{eqfree}), there are only two terms which can vary with system size: the dyon interaction term $\Delta f$ and the third term stemming from Stirling's approximation, which has explicit $\tilde{V}_3$-dependence. Increasing the system size decreases the contribution from the third term as it goes as $1/\tilde{V}_3$. Subtracting this term from the results reveals that the contribution from the dyon interactions increases with system size. Both of these effects are reduced in the deconfined phase where the density is lower.
    
    The main factor in the finite-size effects regarding dyon interactions comes from dyons near the boundaries of the system. As long as a finite number of dyons and image boxes are used, there are dyons near the faces of the cube whose total short-range interactions are reduced. The fraction of dyons near the faces of the cube should scale as $N_D^{-1/3}$ and thus vanishes in the thermodynamic limit. Parameters that determine the effective range of interactions such as the core sizes and Debye mass also play an important role in determining how quickly the results converge to the $N_D \rightarrow \infty$ limit.  
    
   Different quantities may converge at different speeds. From Fig. \ref{fig:size} it is seen that the value of the free energy minimum decreases by $\sim 2\%$ when doubling the number of dyons. A simple linear fit of $f$ as a function of $1/N_D$ gives an infinite-volume value of $f=-4.519$, a $4.5\%$ decrease from the $N_D = 120$ value.  The input parameters of the ensemble - the holonomy and densities - depend only on the \textit{location} of the minimum, which clearly varies much less. Fitting the curves to quadratic forms gives the densities $n_M = 0.607$, $0.609$, $0.607$ for dyon numbers $N_D= 120$, $180$, $240$, respectively. These values agree within uncertainty and show no clear trend. Similar behavior was seen for the $SU(2)$ results \cite{Lopez-Ruiz:2016bjl}, which studied finite volume effects in more detail and showed good results for a maximum ensemble size of $N_D = 88$. We conclude that the number of dyons $N_D= 120$ used in this work is sufficiently large to determine the thermodynamic properties of the dyon ensemble.
 
   \section{Topological charge fluctuations}
 Here we present some general discussion of topological susceptibility measurements in our setting and on the lattice. 
 
   Suppose a lattice of 4-volume $V_4$ is used, and the 
 instantons with average density $d_I$ populate it $randomly$.
 This leads to Poisson distribution with mean number 
 $\langle n_I \rangle=  d_I V_4$. The same is assumed for antiinstantons $\bar I$, and, as is well known, the susceptibility is simply the total density
  \begin{equation}
  \chi={\langle (n_I-n_{\bar I})^2  \rangle \over V_4}=
  d_I+d_{\bar I}
   \end{equation} 
  
In instanton liquid simulations the numbers $N_I,N_{\bar I}$ are fixed, and fluctuations appear only if
one uses the sub-box as proposed in Ref. \cite{Shuryak:1994rr} for studies of 
correlations in the instanton ensemble. 
Let the sub-box
have volume fraction $f\equiv v_4/V_4$. So, if dyons are
also placed $randomly$, their distribution is binomial 
\begin{equation} 
B(f,N,n)={N! \over n! (N-n)!} f^n (1-f)^{N-n}    
\end{equation}      
In the limit when all volumes -- and thus numbers involved -- are  large, standard statistical mechanics textbook analysis leads to the conclusion that both Poisson and binomial distributions lead to  Gaussian fluctuations around their corresponding mean.
Standard use of Stirling formula $log(n!)=n log(n/e)$
and $n=f N+\delta$ leads to 
\begin{equation}
B(\delta)\sim exp\big[- { \delta^2 \over 2 f (1-f) N}
\big]\end{equation}
or $\langle \delta^2 \rangle=f (1-f) N$. The topological susceptibility of sub-box is then
\begin{equation}\chi(f)={f (1-f)(N_I+N_{\bar I}) \over f V_4}=(1-f)\chi_P
\end{equation}
where $\chi_P$ is that for unrestricted Poisson distribution defined above. Indeed, the fluctuation must vanish at $f\rightarrow 1$, but  for small subsystem $f\rightarrow 0$ both become the same.
So, in comparing the {\em absolute value} of the topological susceptibility we measure to those on the lattice, one need to multiply the former by $1/(1-f)=2$.
  
  Let us now calculate $\chi$ for randomly placed instanton-dyons. Their total numbers 
  in our setting are also fixed, and we always keep $$N_L=N_{\bar L},\,\,\, N_{M1}=N_{\bar M1} ,
\,\,\, N_{M2}=N_{\bar M2} $$   
which ensures that the total magnetic and topological charges of the whole system to be zero. 
The topological charge in sub-box is
     \begin{equation} 
Q=(1-2\nu)(n_L-n_{\bar L})+ \nu(n_{M1}-n_{\bar M1})+\nu(n_{M2}-n_{\bar M2})
   \end{equation}      	
It is integer  in two limiting cases, confining holonomy $\nu=1/3$ at $T<T_c$,  and trivial holonomy, $\nu=0$ at large $T$. The unrestricted topological susceptibility is then
 \begin{equation} 
 \chi=\frac{\big((1-2\nu)^2 (N_L+N_{\bar L})+\nu^2(N_{M1}+N_{M2}+N_{\bar M1}+N_{\bar M2}) \big)}{V_4}
 \end{equation}    	
In the confining case, with $\nu=1/3$ and all dyon numbers equal $N_d$, it is $6N_d/9 V_4$. Comparing it to the case
in which any triplet $L,M_1,M_2$ is tightly coupled into instantons with the same density $N_I=N_d$, one finds that
 \begin{equation}
 \chi(random \,dyons)={1 \over 3} \chi(fully \, correlated \,dyons)
 \end{equation}  
Indeed, the clustering of dyons into instantons increases the fluctuations. Conversely, the clustering of dyons into $D\bar D$ pairs (as indeed happens at higher temperatures) decreases the fluctuations.

   \onecolumngrid

  \bibliography{MD_dyons1}

\end{document}